\documentclass{article}
\usepackage{graphicx}
\usepackage{cite}
\usepackage{epsfig}
\usepackage{amsfonts}
 \usepackage{amssymb}
\usepackage{amsmath}
\usepackage{slashed}
\usepackage{soul}
\usepackage{hyperref}
\usepackage{dcolumn}        
\usepackage{bm}            		 
\usepackage{amstext}
\usepackage[dvipsnames]{xcolor}
\usepackage{color} 
\usepackage{transparent}
\usepackage{rotating}
\usepackage{array}
\usepackage[T1]{fontenc}
\usepackage[utf8]{inputenc}
\newcommand\mybar{\kern1pt\rule[-\dp\strutbox]{.8pt}{\baselineskip}\kern1pt}
\usepackage{float}

\newcolumntype{P}[1]{>{\centering\arraybackslash}p{#1}}

\date{\today}

\begin{document}
\begin{center}

{\Large \bf Virial  identities in relativistic gravity: \\ 1D effective actions and the role of boundary terms}
\vspace{0.8cm}
\\
{Carlos A. R. Herdeiro$^{\ddagger}$, Jo\~ao M. S. Oliveira$^{\dagger}$, 
Alexandre M. Pombo$^{\ddagger}$, Eugen Radu$^{\ddagger}$,  \\
\vspace{0.3cm}
$^{\ddagger }${\small Departamento de Matem\'atica da Universidade de Aveiro and } \\ {\small  Centre for Research and Development  in Mathematics and Applications (CIDMA),} \\ {\small    Campus de Santiago, 3810-183 Aveiro, Portugal}
\\
\vspace{0.3cm}
$^{\dagger}${\small Centro de Astrof\'\i sica e Gravita\c c\~ao - CENTRA,} \\ {\small Departamento de F\'\i sica,
Instituto Superior T\'ecnico - IST, Universidade de Lisboa - UL,} \\ {\small Avenida
Rovisco Pais 1, 1049-001 Lisboa, Portugal}
\vspace{0.3cm}
}
\end{center}

	\begin{abstract}   
Virial (\textit{aka} scaling) identities are integral identities that are useful for a variety of purposes in non-linear field theories, including establishing no-go theorems for solitonic and black hole solutions, as well as for checking the accuracy of numerical solutions. In this paper, we provide a pedagogical rationale for the derivation of such integral identities, starting from the standard variational treatment of particle mechanics. In the framework of one-dimensional (1D) effective actions, the treatment presented here yields a set of useful formulas for computing virial identities in any field theory. Then, we propose that a complete treatment of virial identities in relativistic gravity must take into account the appropriate boundary term. For General Relativity this is the Gibbons-Hawking-York boundary term. We test and confirm this proposal with concrete examples. Our analysis here is restricted to spherically symmetric configurations, which yield 1D effective actions (leaving higher-D effective actions and in particular the axially symmetric case to a companion paper). In this case, we  show that there is a particular "gauge" choice, $i.e.$ a choice of coordinates and parameterizing metric functions, that simplifies the computation of virial identities in General Relativity, making both the Einstein-Hilbert action and the Gibbons-Hawking-York boundary term non-contributing. Under this choice, the virial identity results \textit{exclusively} from the matter action. For generic "gauge" choices, however, this is not the case. 
	\end{abstract}

\medskip

\tableofcontents

%
\section{Introduction}\label{S1}
%

In particle mechanics the virial theorem is a \textit{statistical} result. It provides a useful relation between the averages over time of the total kinetic and potential energies for a stable system of $N$ bound particles. The theorem reads \cite{goldestein} 
\begin{equation}
\langle T \rangle =-\frac{1}{2}\sum_{i=1}^N \langle \vec{F}_i\cdot \vec{r}_i \rangle \ ,
\label{virial1870}
\end{equation}
where $T$ denotes the total kinetic energy and $\vec{F}_i$ the force over the $i^{\rm th}$ particle, which has position $\vec{r}_i$. The time averaging, denoted by $\langle \rangle$, amounts to a time integral, $\langle A \rangle \equiv (\Delta t)^{-1}\int_{t_i}^{t_f}A\, dt$, for  any quantity $A$.  Upon choosing appropriately an  integration interval $\Delta t\equiv t_f-t_i$, the theorem is, equivalently,
\begin{equation}
\frac{1}{\Delta t}\int_{t_i}^{t_f} \left( T + \frac{1}{2}\sum_{i=1}^N  \vec{F}_i\cdot \vec{r}_i \right)dt=0 \ .   \qquad \qquad {\rm {\bf [virial \  Clausius]}}
\label{virial18702}
\end{equation}

Eq.~\eqref{virial18702} makes clear that the virial theorem amounts to an \textit{integral identity}. If the motion is periodic, choosing $\Delta t$ to be a multiple of the period, the integral exactly vanishes and the $(\Delta t)^{-1}$ pre-factor is unnecessary. But even if the time integration is not exactly zero (for instance if the motion is not periodic), for a system of bound stable particles, the integrand is bounded, and the \textit{lhs} of~\eqref{virial18702} can be made arbitrarily small choosing a sufficiently large time interval. In either case, the virial theorem holds to arbitrary accuracy.

If the forces are conservative, derivable from a total potential energy $U$, and if $U$ is a homogeneous function of degree $n$ of the particles' coordinates, then the virial theorem takes the form $\langle T\rangle =n \langle U\rangle/2$~\cite{goldestein}.
For the special case of inverse square law forces, $n=-1$, we recover the familiar result that the average kinetic energy (in modulus) is one half of the average potential energy (which is negative):\footnote{Here the integral is understood to be over a multiple of the period.} 
\begin{equation}
\int_{t_i}^{t_f} \left( T + \frac{U}{2}   \right)dt=0 \ .    \qquad \qquad {\rm {\bf [virial \  inverse \ square \ force \ law]}}
\label{virialkepler}
\end{equation} 

The virial identity~\eqref{virialkepler} can be recovered by a \textit{scaling argument}. Consider the classical action of a particle, $\mathcal{S}=\int_{t_i}^{t_f}(T-U)dt$, where the kinetic energy is $T$ (which is a homogeneous function of degree 2 of the velocity) and the potential energy is $U=U(\vec{r})$, here assumed to be a homogeneous function of $\vec{r}$ of degree $n$. Consider that there is a solution of the classical equations of motion $\vec{r}=\vec{r}(t)$.
	If one \textit{scales} this fiducial solution by a factor of $\alpha$,  $\vec{r}(t)\rightarrow \alpha \vec{r}(t)$, then $T\rightarrow \alpha ^2 T$, while $U\rightarrow  \alpha ^nU$. The corresponding action\footnote{The action of the scaled solution becomes a \textit{function} of $\alpha$, whereas it is a \textit{functional} of the particle's path.} $\mathcal{S}_\alpha=\int_{t_i}^{t_f}(\alpha^2 T-\alpha^n U)dt$ should be stationary at the original fiducial solution:
\begin{equation}
\frac{\partial \mathcal{S}_\alpha}{\partial \alpha}\bigg|_{\alpha=1}= 0 \ \ \stackrel{n=-1 \ , \  \Delta t = {\rm period}}{\Rightarrow} \ \ \eqref{virialkepler} \ .
\end{equation}

Note that $n=-1$ guarantees the motion is periodic and choosing $\Delta t$=\rm period makes the above scaling a variational problem with periodic boundary conditions rather than fixed boundary conditions. This illustrates the derivation of a virial identity from a scaling argument. 

\medskip

Originally presented by R. Clausius in 1870~\cite{doi:10.1080/14786447008640370}, who dubbed the \textit{rhs} of~\eqref{virial1870} "virial", the virial theorem has found many applications in physics and mathematics. In the context of gravitation, for instance, F. Zwicky first deduced the existence of a gravitational anomaly, and suggested the existence of  "dark matter", from an application of the virial theorem~\cite{Zwicky:1933gu}.

 In this paper we shall be interested in integral identities that are virial-like (and thus, following the literature,  will be referred to as "virial identities"), but in field theory rather than particle mechanics, obtained from \textit{scaling} arguments.
The first example of such virial identities in field theory arose as a "no-go" theorem for \textit{solitons}. 

The possible existence of soliton-type configurations (particle-like solutions inspired by \textit{solitary} wave solutions of the Korteweg-de-Vries equation~\cite{korteweg1895xli,whitham1965non,darrigol2005worlds,dauxois2006physics}) emerges as an interesting question in any non-linear field theory. The robustness against decay of the `shape' of such solutions is interpreted as a cancellation between non-linear and dispersive effects. In this context, \textit{Derrick's theorem}~\cite{derrick1964comments} was put forward in 1964 as a generic argument against the existence of stable, finite energy, time-independent solutions in a wide class of non-linear wave equations, in three or higher (spatial) dimensions - see also~\cite{hobart1963instability,hobart1965non} for an earlier similar argument. This theorem results from a scaling argument; for a 1+3 dimensional relativistic scalar field theory of a scalar field $\Phi$, with spatial gradiant $\nabla \Phi$ and potential energy $U(\Phi)$, it results in the virial identity ($cf.$ Section~\ref{S31})

\begin{equation}\label{virialderrick}
	 \int d^3{\bf r} \left[\frac{(\nabla \Phi)^2}{3}+U(\Phi)\right] =0\ . \qquad \qquad {\rm {\bf [virial \  Derrick]}}
	\end{equation}
Eq.~\eqref{virialderrick} represents the prototypical virial identity in field theory. It has a simple interpretation. If the potential energy is non-negative, since $(\nabla \Phi)^2\geqslant 0$, then~\eqref{virialderrick} can only be obeyed for a constant $\Phi=\Phi_0$ (for which $U(\Phi_0)=0$). Thus, there are no non-constant configurations, hence no solitons. 

The usefulness of virial identities is not exhausted in establishing no-soliton theorems. In generic setups, which includes more general field theories (possibly also with gravity) and more general ansatze  for the fields, virial/scaling identities serve to understand the balance between the different effects that allow the existence of solitonic or black hole (BH) solutions (see $e.g.$ Section~\ref{S32}). In this sense, virial identities serve as a guide to construct new solutions. Additionally, as for solitons, they can also be used to establish no-go theorems for BHs with non-trivial matter fields, also known as "no-hair" theorems - see $e.g.$~\cite{heusler1992scaling,heusler1996no,Herdeiro:2015waa}. Furthermore, in the context of numerical solutions, virial identities serve as useful identities to test the accuracy of such numerical solutions - see $e.g.$~\cite{herdeiro2018spontaneous,Fernandes:2019rez,Fernandes:2019kmh}.

Despite these (and other) interesting applications, the use of virial identities in the context of strong gravity as been mostly restricted to spherically symmetric solutions and a particular "gauge" (by which we mean a coordinates \textit{plus} a parameterization) choice. The main goal of this paper is to present a generic methodology for establishing virial identities for equilibrium, asymptotically flat, localized configurations using any "gauge" choice for the metric and matter fields. In doing so, we will unveil a key ingredient, hitherto neglected, that must be taken into account in relativistic gravity applications - in general, there can be a non-trivial contribution from boundary terms. In the context of General Relativity (GR) the appropriate boundary term is the Gibbons-Hawking-York (GHY) term~\cite{york1972role,gibbons1993action}, which must be considered in order to derive the correct virial identity.

After establishing a general methodology, we shall test the so obtained virial identities, providing examples corresponding to different field theories and parameterization choices. One can face the virial identity in a certain model, encompassing different fields as a "word" composed by different "letters". Computing the basic "letters" one can efficiently piece them together into the virial identity "word", for a model composed by the different fields analysed here.  Moreover, our analysis reveals a simpler "gauge" choice for which the gravitational part does not contribute. There is, therefore, a simple setup to compute virial identities in GR just by computing the contribution of the matter action, which can be safely used by virtue of the generic understanding presented here. 

This paper is focused on spherically symmetric configurations, leaving the treatment of axially symmetric configurations to a companion paper~\cite{companion}.  It is organised as follows. We start in Section~\ref{Section2} by considering the variational treatment in particle mechanics. This Section serves two purposes. Firstly it builds a bridge between the scaling transformation that yields virial identities and the familiar standard variational treatment in Lagrangian mechanics. Secondly, it introduces the notion of \textit{effective action} (EA) that, in practice, is the central object used in building virial identities in field theory that yield a 1D EA (as in spherical symmetry). As we shall see, the virial identities obtained in this Section (eqs.~\eqref{virialea}, \eqref{virialea2}, \eqref{virialea3} and \eqref{virialea4}) can then be used as general formulae for the subsequent problems found in field theory. In Section~\ref{Section3}, we review Derrick's theorem as the paradigmatical illustration of a scaling argument and of a virial identity. But we also show how a change of ansatz leads to a way of circumventing Derrick's theorem allowing the existence of scalar field theory, flat spacetime solitons known as $Q$-balls~\cite{Coleman:1985ki}. In Section~\ref{Section4}, we take a first look at GR. This section is meant as pedagogical, and the virial relations obtained therein are \textit{incomplete}. Our goal is to illustrate two points. Firstly, there are simpler "gauge" choices to compute virial identities. In the simplest parameterization, the Einstein-Hilbert (EH) action results in a scale invariant EA; then it does not contribute to the virial identity. Secondly, by considering the case of electrovacuum, we show that the (would be) virial identity derived solely from the EH-Maxwell action is not correct, as it is not obeyed by the Reissner-Nordstr\"om (RN) solution. The complete treatment is then introduced in Section~\ref{Section5}, where we include the contribution of the GHY boundary term and we provide the complete virial identities for the vacuum and electrovacuum cases. In Section~\ref{Section6}, we take advantage of the simplest "gauge" choice to compute the virial identity for various examples of field theories minimally coupled to Einstein's gravity, by considering simply the contribution of the matter part. To emphasise the generic case, however, in Section~\ref{Section7} we discuss the virial identities for electrovacuum and (massive-complex) scalar-vacuum in isotropic coordinates, confirming the non-trivial contribution from the gravitational part, that is mandatory in order for the virial identity to be obeyed by known solutions. We provide a discussion and our conclusions in Section~\ref{Section8}. In this paper we use units with $G=1=c$.

\section{Particle mechanics and effective actions}
\label{Section2}

Some insight and useful formulas that  will be used in the field theory case can be obtained by addressing first particle mechanics. Let us start with a recap of the elementary variational treatment. 

\subsection{The standard variational treatment}
\label{S21}

Consider an action functional $\mathcal{S}$, depending on a set of $n$ generalized coordinates $q_j$ ($j=1\dots n$  ), their first time derivatives, $\dot{q}_j$, and on the time coordinate $t$ (so that $\dot{q}_j = dq_j/dt$). The action is the time integral of the Lagrangian $L$:
	\begin{equation}
	 \mathcal{S}[q_j(t),\dot{q}_j(t),t] = \int_{t_i}^{t_f}L\left(q_j,\dot{q}_j,t\right)dt \ .
	\end{equation}
In the standard variational problem one aims at finding the true \textit{path} of the particle in $\mathbb{R}^{n}$, which is a map
\begin{align}
[t_i,t_f]\in \mathbb{R} & \rightarrow \mathbb{R}^{n} \nonumber \\
t  & \rightarrow q_j (t) \ ,
\end{align}
traveled as a function of (time) $t$. This  path extremises the action functional. To compute it, one considers an arbitrary variation $\delta q_j(t)$ around a fiducial path, $q_j(t)$, where the endpoints are fixed, $\delta q_j(t_i)= \delta q_j(t_f)=0$. This generates a variation of the action $\delta \mathcal{S}$. Hamilton's principle (\textit{aka} principle of least action) selects the true path as the fiducial path if $\delta \mathcal{S}\Big|_{\delta q_j=0}=0$. 

Explicitly, the variation (using the chain rule and integrating by parts) reads
	\begin{align}
	 \delta \mathcal{S} 
= \int_{t_i}^{t_f} \delta L dt 
= \int_{t_i}^{t_f}\left( \frac{\partial L}{\partial \dot{q}_j}\delta\dot{q}_j +  \frac{\partial L}{\partial q_j}\delta q_j + \frac{\partial L}{\partial t}\delta t\right)dt
	 &=\left[\frac{\partial L}{\partial \dot{q}_j}\delta q_j\right|^{t=t_f}_{t=t_i} + \int_{t_i}^{t_f}\left[ -\frac{d}{dt}\left(\frac{\partial L}{\partial \dot{q}_j}\right) +  \frac{\partial L}{\partial q_j}\right]\delta q_jdt  \ .
\label{standardvariation}
	\end{align}
For arbitrary variations under fixed endpoints, the first term of the \textit{rhs} of the last equation vanishes, and the second terms yields a set of  \textit{differential} requirements for the true path, the 
\textit{Euler-Lagrange equations}
	\begin{equation}
	 \frac{d}{dt}\left(\frac{\partial L}{\partial \dot{q}_j}\right) = \frac{\partial L}{\partial q_j} \ .
	\end{equation}

%
\subsection{A scaling transformation of an effective action}
\label{S22}
%
In the standard variational treatment~\eqref{standardvariation} the term $(\partial L/\partial t) \delta t$ was dropped under the assumption that the Lagrangian has no explicit dependence on $t$. Moreover, arbitrary variations of the path were considered. We shall now consider a variation on the variational problem, where an explicit dependence on (the analogue of) $t$ is present and it is a variation of this parameter that induces the variation of the "path". Instead of considering the \textit{path} traveled in time by a particle in $\mathbb{R}^{n}$, however, we shall consider the (spatial) \textit{profile} of a map:
\begin{align}
[r_i,\infty]\in \mathbb{R} & \rightarrow \mathbb{R}^{n} \nonumber \\
r  & \rightarrow q_j (r) \ ,
\end{align}
which is spanned as a function of a (spatial) coordinate $r$. Having in view the field theory applications below, we choose the profile to start at $r=r_i$ and end at $r=+\infty$. There are infinitely many possible profiles, but the true one extremizes a certain \textit{effective action} (EA)
	\begin{equation}
	 \mathcal{S}^{\rm eff}[q_j(r),q'_j(r),r] = \int_{r_i}^{\infty}\mathcal{L}\left( q_j,q'_j,r\right)dr \ ,
\label{actionspatial}
	\end{equation}
where $q'_j(r)=dq_j(r)/dr$. This EA does not have the physical dimensions of an action. But it plays the role of an action in the sense that it determines  the true configurations through a variational principle. By the same token we shall be referring to the integrand in~\eqref{actionspatial} $\mathcal{L}$ as an effective Lagrangian. 

In the standard variational treatment, we have considered arbitrary variations of a fiducial path $q_j(t)$. Now, we shall vary the independent parameter $r$ in a specific manner, and consider the profile variation induced by the latter. 
Concretely,  we consider a transformation $r\rightarrow \tilde{r}$ that \textit{scales} $r$ but keeps $r_i$ as a fixed point.  Thus
	\begin{equation}
	 r\rightarrow \tilde{r} = r_i + \lambda(r-r_i) \ ,
\label{scaling1}
	\end{equation}
where $\lambda$ is an arbitrary positive constant, 	such that  $\tilde{r}=r_i$ for $r=r_i, \forall_\lambda$ (fixed point); the transformation trivializes for $\lambda=1$:   $\tilde{r} = r $. The new profile  induced by the scaling~\eqref{scaling1} is 
\begin{equation}
q_j(r)\rightarrow q_{\lambda j}(r)=q_j( \tilde{r}) \ .
\end{equation} 
The EA of the scaled profile becomes a \textit{function} of $\lambda$, denoted as $\mathcal{S}_\lambda^{\rm eff}$, 
\begin{equation}
 \mathcal{S}_\lambda ^{\rm eff}
= \int_{r_i}^{\infty}\mathcal{L}_\lambda \left(q_j({r}),\frac{dq_j(r)}{dr},{r}\right)dr 
= \int_{r_i}^{\infty}\mathcal{L}\left(q_j(\tilde{r}),\frac{dq_j(\tilde{r})}{dr},r\right)dr
= \int_{r_i}^{\infty}\mathcal{L}\left(q_j(\tilde{r}),\lambda \frac{dq_j(\tilde{r})}{d\tilde{r}},\frac{\tilde{r}-r_i}{\lambda}+r_i\right)\frac{d\tilde{r}}{\lambda} \ .
\label{sprofile}
\end{equation}
The true profile obeys the \textit{stationarity condition}
	\begin{equation}
	 \frac{\partial \mathcal{S}^{\rm eff}_\lambda}{\partial \lambda} \bigg|_{\lambda=1}=0 \ ,
\label{variationscale}
	\end{equation}
which, from the last equality in~\eqref{sprofile}  yields
\begin{equation}
%
\int_{r_i}^\infty\left[ \sum_j \frac{\partial \mathcal{L}}{\partial q'_j} q'_j -\mathcal{L} -\frac{\partial \mathcal{L}}{\partial r}(r-r_i)\right]dr = 0 \ .  \qquad \qquad {\rm {\bf [virial \  EA \ 1]}}
\label{virialea}
%
\end{equation}

Unlike the standard variational procedure, yielding a set of differential constraints, here we obtain an \textit{integral} constraint that should be obeyed if the $q_j(r)$ are solutions of the Euler-Lagrange equations derived from~\eqref{actionspatial}. Observe that the first two terms in the integrand of~\eqref{virialea} combine into a ``Hamiltonian''
\begin{equation}
\mathcal{H}\equiv  \sum_j \frac{\partial \mathcal{L}}{\partial q'_j} q'_j -\mathcal{L} \ .
\end{equation}

\subsection{Effective Lagrangians depending on second order derivatives}
\label{S23}
%
In field theory, we shall sometimes find effective Lagrangians depending \textit{also} on the second derivative of the profile functions $q''_j(r)=d^2q_j(r)/dr^2$. For instance, the EH Lagrangian ($cf.$~eq.~\eqref{ehaction} below) depends on the second derivatives of the metric. In such cases, to consider the variational problem, the action~\eqref{actionspatial} is replaced by the more general
	\begin{equation}
	 \mathcal{S}^{\rm eff}[q_j(r),q'_j(r),q''_j(r),r] = \int_{r_i}^{\infty}\mathcal{L}\left( q_j,q'_j,q''_j,r\right)dr \ .
\label{actionspatial2}
	\end{equation}
Repeating the procedure of the previous sub-section, \textit{mutatis mutandis}, we obtain the more general virial identity
\begin{equation}
\int_{r_i}^\infty\left[ \sum_j \frac{\partial \mathcal{L}}{\partial q'_j} q'_j+2 \sum_j \frac{\partial \mathcal{L}}{\partial q''_j} q''_j -\mathcal{L} -\frac{\partial \mathcal{L}}{\partial r}(r-r_i)\right]dr = 0 \ .  \qquad \qquad  {\rm {\bf [virial \ EA \ 2]}}
\label{virialea2}
\end{equation}

\subsection{Scalings affecting the integration limits}
\label{S24}
A further generalization is to consider a scaling that affects the integration limits. The simplest example is to replace~\eqref{scaling1} by
\begin{equation}
	 r\rightarrow \tilde{r} =  \lambda r \ .
\label{scaling2}
	\end{equation}
This transformation impacts non-trivially on the lower limit of the action integral \eqref{actionspatial2}. To understand the corresponding contribution to the virial identity, we repeat the steps in eq.~\eqref{sprofile} (allowing, as in Section~\ref{S23}, a further $q''_j(r)$ dependence) to find
\begin{equation}
 \mathcal{S}^{\rm eff}_\lambda 
= \int_{\lambda r_i}^{\infty}\mathcal{L}\left(q_j(\tilde{r}),\lambda \frac{dq_j(\tilde{r})}{d\tilde{r}},\lambda^2 \frac{d^2q_j(\tilde{r})}{d^2\tilde{r}},\frac{\tilde{r}}{\lambda}\right)\frac{d\tilde{r}}{\lambda} \ .
\label{sprofile2}
\end{equation}
Thus, the stationarity condition \eqref{variationscale} now yields an extra term:
\begin{equation}
\int_{r_i}^\infty\left[ \sum_j \frac{\partial \mathcal{L}}{\partial q'_j} q'_j+2 \sum_j \frac{\partial \mathcal{L}}{\partial q''_j} q''_j -\mathcal{L} -\frac{\partial \mathcal{L}}{\partial r}r\right]dr = r_i \mathcal{L}(r_i) \ . \qquad \qquad  {\rm {\bf [virial \ EA \  3]}}
\label{virialea3}
\end{equation}

%
\subsection{Adding a total derivative to the effective Lagrangian}
\label{S25}
%
As a final discussion point, leading in fact to the formula that will be most used in the field theory applications below, we observe that in some circumstances there are boundary terms that can be added to the Lagrangian, which take the form of a total derivative. Consequently, these terms do not affect the bulk equations of motion. A total derivative can, however, affect the virial identity. Typically there can be a trade off between considering a total derivative \textit{or} considering an effective Lagrangian with second order derivatives (as in Section~\ref{S23}). The virial identities obtained using either perspective are equivalent (for an illustration see Section~\ref{S41} below).

To see the explicit form of the virial identity when a total derivative is present, consider an EA\footnote{When considering a total derivative we do not consider second derivatives in the effective Lagrangian, due to the trade off between these two types of terms.}
	\begin{equation}
	 \mathcal{S}^{\rm eff}[q_j(r),q'_j(r),r] = \int_{r_i}^{\infty}\hat{\mathcal{L}}\left( q_j,q'_j,r\right)dr \ ,
\label{actionspatial3}
	\end{equation}
where the new Lagrangian $\hat{\mathcal{L}}$ contains a total derivative term
\begin{equation}
\hat{\mathcal{L}}\left(q_i,q'_i,r\right)=\mathcal{L}\left(q_i,q'_i,r\right)+\frac{d}{dr}f\left(q_i,q'_i,r\right) \ ,
\end{equation}
and $f$ is some function that depends on the same variables as the original effective Lagrangian $\mathcal{L}$, up to first derivatives. Performing the scaling~\eqref{scaling1}, the stationarity condition \eqref{variationscale} now yields
\begin{equation}
%
\int_{r_i}^\infty\left[ \sum_j \frac{\partial \mathcal{L}}{\partial q'_j} q'_j -\mathcal{L} -\frac{\partial \mathcal{L}}{\partial r}(r-r_i)\right]dr = \left[\frac{\partial f}{\partial r}(r-r_i) - \sum_i\frac{\partial f}{\partial q'_i} q'_i\right]^{+\infty}_{r_i} \ .  \qquad \qquad   {\rm {\bf [virial \ EA \  4]}}
\label{virialea4}
%
\end{equation}

Eqs.~\eqref{virialea}, \eqref{virialea2}, \eqref{virialea3} and \eqref{virialea4} provide useful relations that can be readily used in the context of EAs obtained from field theory models, as illustrated in the next Sections.

\section{Flat spacetime field theory}\label{Section3}
Let us now address two examples in flat spacetime relativistic (scalar) field theory. The mandatory first example is to review the original theorem by Derrick~\cite{derrick1964comments}, establishing the inexistence of solitions in a large class of non-linear field theories. We then consider a more generic ansatz for the scalar field configuration (allowing a harmonic time-dependence) and illustrate how the virial identity is compatible with the existence of solitons known as \textit{$Q$-balls}~\cite{Coleman:1985ki}.

\subsection{Derrick's theorem}\label{S31}
%

Consider the (possibly) non-linear Klein-Gordon equation, describing a real scalar test field on Minkowski spacetime:
	\begin{equation}
	 \Box\Phi=\frac{1}{2}\frac{dU}{d\Phi} \ ,
	\end{equation}
	where $U(\Phi)$ is a potential energy function.  This can be derived from the following "matter" action:
	\begin{equation}\label{realaction}
	 \mathcal{S}_{\rm m}^\Phi=\frac{1}{4\pi}\int d^4x\left[-\partial_\mu \Phi \partial^\mu \Phi -U(\Phi)\right] \ .
	\end{equation}
	Splitting the spacetime coordinates $x^\mu=(t,{\bf r})$ into temporal and spatial coordinates, the action may be rewritten as:
	\begin{equation}
	 \mathcal{S}_{\rm m}^\Phi=\frac{1}{4\pi}\int dt \left( S_0-S_1-S_2 \right) \ ,
	\end{equation}
	where
	\begin{equation}
	 S_0\equiv \int d^3{\bf r} (\partial_t \Phi)^2 \ , \qquad S_1\equiv \int d^3{\bf r} (\nabla \Phi)^2 \ , \qquad S_2\equiv \int d^3{\bf r} U(\Phi) \ ,
	\end{equation}
	and the integration is over the whole space. We will prove that no stable, time-independent, localised solutions exist, for any potential energy. Time-independence implies $S_0=0$. By \textit{localized} we mean that $S_1$ and $S_2$ are finite. Due to the time-independence we may consider the EA
	\begin{equation}
	 \mathcal{S}^{\rm eff}=S_1+S_2 \ .
	\end{equation}
	The existence of a localized solution, by Hamilton's principle, implies $\delta  \mathcal{S}^{\rm eff}=0$. Let the solution be $\Phi({\bf r})$; due to the time-independence, extremizing the EA is equivalent to extremizing the energy ($\delta  \mathcal{S}^{\rm eff} = \delta E$). The solution is stable if $\delta^2E\geqslant 0$.
	
	Let us define a scaled configuration $\Phi_\lambda({\bf r})=\Phi(\lambda {\bf r})$, where  the radial coordinate suffers the dilation $r\rightarrow \tilde{r}=\lambda r$. The energy of such scaled configuration is:
	\begin{equation}
	 E_\lambda= \int d^3{\bf r}\left[ (\nabla \Phi_\lambda)^2 +U(\Phi_\lambda)\right] =\frac{S_1}{\lambda}+\frac{S_2}{\lambda^3} \ .
	\end{equation}
	Since, by assumption, the original configuration $\Phi({\bf r})$ (corresponding to $\lambda =1$) was a solution 
	\begin{equation}\label{virial0}
	 \left(\frac{dE_\lambda}{d\lambda}\right)_{\lambda=1}=-S_1-3S_2=0 \ , \qquad \Leftrightarrow \qquad {S_2}=-\frac{S_1}{3} \ .
\end{equation}
Equation~(\ref{virial0}) is Derrick's virial identity, eq.~\eqref{virialderrick}. It relates the total "kinetic" and potential energy.
As mentioned in the Introduction, inspection thereof is physically insightful: since the first term in the square bracket is clearly everywhere positive, for positive definite potentials there can be \textit{no solution}, regardless of being stable or not.  On the other hand, 
	\begin{equation}
	 \left(\frac{d^2E_\lambda}{d\lambda^2}\right)_{\lambda=1}=2S_1+12S_2\stackrel{~\eqref{virial0}}{=}-2S_1<0 \ ,
	\end{equation}
	since $S_1$ is manifestly positive. It follows that for \textit{any} $U$, even if it allows the existence of a solution (which may be the case for a non-positive $U$), the stretching of the hypothetical solution decreases its energy and thus, such a solution is unstable. These arguments illustrate how virial identities can establish \textit{no-go} theorems.  A straightforward generalization to higher dimensions can be found in Appendix~\ref{AppendixA}.

\subsection{Circumventing Derrick's theorem: $Q$-balls}
\label{S32}
%
	 In the original work~\cite{derrick1964comments}, Derrick observed that one way to circumvent the theorem would be to allow localized solutions that are periodic in time, rather than time-independent. For a real scalar field, however, such configuration would not be static (or stationary). Various authors, starting with
Rosen~\cite{rosen1968particlelike}, considered a \textit{complex scalar field} $\Phi$, described by the matter action\footnote{Here `$^*$' denotes complex conjugate and, albeit still in flat spacetime, we allow the Minkowski metric $g$ to be written in curvilinear coordinates. 
}
	\begin{equation}\label{complexaction}
	 \mathcal{S}_{\rm m}^{\Phi^*} =\frac{1}{4\pi}\int d^4x \sqrt{-g}  \bigg[-\frac{1}{2}g^{\mu\nu}(\partial_\mu\Phi\partial_\nu\Phi^* + 	\partial_\mu\Phi^*\partial_\nu\Phi)-U(|\Phi|)\bigg] \ ,
	\end{equation}
	with a harmonic time-dependence:
	\begin{equation}\label{complexPhi}
	 \Phi(t,r)=\phi(r) e^{-i\omega t} \ ,
	\end{equation}
	which guarantees a time-independent energy-momentum tensor. Moreover, there is a global symmetry and a conserved scalar Noether charge. Then, for some classes of potentials (yielding non-linear models), localized stable solutions exist, which are known, following Coleman~\cite{Coleman:1985ki}, as $Q$-balls (since the Noether charge is typically labelled $Q$).

Let us derive a virial identity for spherical solutions in this model, to analyse how it is compatible with the existence of spherical $Q$-balls.  We  use the standard spatial spherical  coordinates for the Minkowski background: $(t,r,\theta,\phi)$.
	Due to the spherical symmetry, the action is $(\theta,\varphi)$-independent and these terms can be integrated right away. Repeating Derrick's argument, we now have that $\mathcal{S}_{\rm m}^{\Phi^*} =- \int dt \mathcal{S}^{\rm eff}$, where the EA $\mathcal{S}^{\rm eff}$ is written as:
	\begin{equation}
	 \mathcal{S}^{\rm eff}= \int_0^\infty  dr \, r^2 \left[- \omega^2\phi^2+\left(\frac{d\phi}{dr}\right)^2 +U(|\phi|)\right]\equiv S_0+S_1+S_2 \ .
\label{effqballs}
	\end{equation}
	Consider, again, a rescaled configuration $ \phi_\lambda(r)=\phi(\lambda r)$. Its EA is 
	\begin{equation}
	 \mathcal{S}^{\rm eff}_ \lambda=  \int_0^\infty dr  \, r^2 \left[ -\omega^2\phi_\lambda^2+\left(\frac{d\phi_\lambda}{dr}\right)^2 +U(|\phi_\lambda|)\right]=\frac{S_0+S_2}{\lambda^3} + \frac{S_1}{\lambda}\ .
	\end{equation}
	Thus
	\begin{equation} \label{virial02}
	 \left(\frac{d\mathcal{S}^{\rm eff}_ \lambda}{d\lambda}\right)_{\lambda=1}=0 \  \qquad \Leftrightarrow \qquad {S_0+S_2}=-\frac{S_1}{3} \ ,
	\end{equation}
	or, explicitly, 
	\begin{equation} \label{virialqballs}
	 \int_0^\infty  dr \,  r^2 \left[- \omega^2\phi^2+\frac{1}{3}\left(\frac{d\phi}{dr}\right)^2 +U(|\phi|)\right] = 0 \ . \qquad \qquad {\rm {\bf [virial  \ {\rm Q-}balls]}}
	\end{equation}
One observes that the harmonic time-dependence yields a term with the opposite sign $(-\omega^2\phi^2)$, so that the obstruction raised by Derrick's theorem does not necessarily apply. 
	The existence of solutions, however, depends on the choice of the potential. If one chooses the potential to be solely a mass term $U(\phi)=\mu^2\phi^2$, then~\eqref{virialqballs} becomes:
	\begin{equation}
	 \int_0^\infty  dr \,  r^2 \left[(\mu^2- \omega^2)\phi^2+\frac{1}{3}\left(\frac{d\phi}{dr}\right)^2 \right] = 0 \ ,
	\end{equation}
	and for bound states, which obey $\omega<\mu$, one immediately concludes the inexistence of solutions. In other words, the virial identity~\eqref{virialqballs} implies that the scalar field must have self-interactions, even with the harmonic time-dependence, in order to yield solitonic solutions. Indeed, $Q$-balls are constructed taking an everywhere positive potential with self-interactions, and for which $U(\phi)-\omega^2\phi^2<0$ in some spatial regions.

Finally, let us remark how~\eqref{virialqballs} can be readily obtained from applying the virial identity formulas for the EAs in Section~\ref{Section2}. Comparing~\eqref{effqballs} with~\eqref{actionspatial} one identifies $r_i=0$ and the effective Lagrangian
\begin{equation}
\mathcal{L}(\phi,\phi',r)=r^2 \left[- \omega^2\phi^2+\left(\phi'\right)^2 +U(|\phi|)\right]\ .
\end{equation}
Then, applying~\eqref{virialea}, a one line computation yields~\eqref{virialqballs}.

\section{GR in spherical symmetry - an incomplete treatment}
\label{Section4}
%
We now consider Einstein's gravity. When deriving solutions of the field equations, one considers the EH action 
\begin{equation}
\mathcal{S}_{\rm EH}=\frac{1}{16\pi}\int d^4x \sqrt{-g} R  \ ,
\label{ehaction}
\end{equation}
where $R$ is the Ricci scalar of the spacetime metric $g_{\mu\nu}$ with determinant $g$. In this way one neglects possible boundary terms. As such, in this Section, we shall be considering models with total action
\begin{equation}
\mathcal{S}=\mathcal{S}_{\rm EH}+ \mathcal{S}_{\rm m} \ ,
\label{totalaction}
\end{equation}
where $\mathcal{S}_{\rm m}$ is some matter/fields action. This treatment will turn out to be incomplete. To be clear, the (would be) virial identities derived in this Section are incomplete (and will be completed in the next Section). The purpose of this Section is twofold. Firstly, it serves as a pedagogical introduction to the need for the GHY boundary term in the derivation of the correct virial identities. Secondly, it serves as an illustration of how the virial identity  derived for any such model depends \textit{both} on the choice of $\mathcal{S}_{\rm m}$ \textit{and} on the parameterization chosen for the metric. We shall now investigate such "gauge" choices, starting with the simplest possible case: a spherically symmetric spacetime in vacuum GR.

\subsection{Vacuum: $\sigma-N$ parameterization in Schwarzschild coordinates}
\label{S41}

An often used ansatz for a spherically symmetric metric spacetime is
	\begin{equation}\label{metricans}
	 ds^2 = -\sigma^2(r)N(r) dt^2 + \frac{dr^2}{N(r)}+ r^2(d\theta^2+\sin^2\theta d\varphi^2) \ .
	\end{equation}
This ansatz uses Schwarzschild-like coordinates, where $r$ is the areal radius, together with parameterizing functions $\sigma(r)$ and $N(r)$. The EH action can then be reexpressed in terms of an EA $\mathcal{S}_{\rm EH}=(4\pi)^{-1}\int dt \mathcal{S}^{\rm eff}$, where
	\begin{align}
	 \mathcal{S}^{\rm eff} = \int dr\  \sigma r^2 R &= -\int \Big\{ r \big[3 r N' \sigma '+2 N \left(r \sigma '' + 2 \sigma '\right)\big]+ \sigma \left(r^2 N''+4 r N'+2 N-2\right)\Big\} dr \ .
\label{eavacuumgr1}
	\end{align}
A distinctive feature is that this action depends on the second derivatives of $\sigma,N$. The second derivative terms can be collected into a \textit{total derivative}, such that this EA is cast in the form~\eqref{actionspatial3} with
	\begin{equation}\label{f1}
\mathcal{L}(\sigma, N; \sigma',N'; r)=-2 \sigma \left(-1+N+rN'\right) \ , \qquad	 f(\sigma, N; \sigma',N'; r) = -2r^2N\sigma'-r^2N'\sigma \ .
	\end{equation}
Admitting the existence of an event horizon, we take $r_i$ in~\eqref{actionspatial3} to be $r_i=r_H$, such that $N(r_H)=0$. Then, the virial identity is readily obtained from~\eqref{virialea4}, yielding 
	\begin{equation}\label{VirialSphAns1}
	 2\int_{r_H}^{\infty}\sigma \left[N-1+(r-r_H)N'\right]dr= \bigg[\left(2rN\sigma'+rN'\sigma\right)(2r_H-r)\bigg]^{+\infty}_{r_H}  \ . 
	\end{equation}
A test on this identity is provided by the Schwarzschild solution, 
	\begin{equation}
	 N(r) = 1-\frac{2M}{r} \ , \qquad\qquad \sigma(r) = 1 \ ,
\label{schwarzschildsol}
	\end{equation}
	with $M$ constant. Indeed, for these choices both sides of~\eqref{VirialSphAns1} give $-4M$. 
Thus, the total derivative term in the EA, albeit not contributing to the equations of motion, gives a non-trivial contribution to the virial identity~\eqref{VirialSphAns1}.

Alternatively, we could have faced the EA~\eqref{eavacuumgr1} as being of the type of~\eqref{actionspatial2} with an effective Lagrangian depending also on second derivatives:
\begin{equation}
\mathcal{L}(\sigma, N; \sigma',N';\sigma'',N''; r)= -  r \left[3 r N' \sigma '+2 N \left(r \sigma '' + 2 \sigma '\right)\right]- \sigma \left(r^2 N''+4 r N'+2 N-2\right) \ .
\end{equation}
Then, applying~\eqref{virialea2} yields an identity that is equivalent to~\eqref{VirialSphAns1}. This illustrates the equivalence observed between the virial identities~\eqref{virialea2} and~\eqref{virialea4} in concrete examples.

Let us emphasise that, despite the apparently non-trivial check provided by the Schwarzschild solution, the (would be) virial identity~\eqref{VirialSphAns1} is \textit{incomplete}. The correct version will be given below in eq.~\eqref{virialvacuumok}.

\subsection{Vacuum: $\sigma-m$ parameterization in Schwarzschild coordinates}
\label{S42}

Virial identities depend not only on the choice of coordinates but also on the choice of metric functions. This is sharply illustrated by reconsidering the metric ansatz of the previous subsection~\eqref{metricans} but with a seemingly innocuous modification: taking as the parameterizing function the Misner-Sharp mass $m(r)$ function~\cite{Misner:1964je}, instead of $N(r)$, given by
	\begin{equation}
	 N(r) = 1-\frac{2m(r)}{r} \ .
\label{mfunction}
	\end{equation}
	In this case, the EA can be written as
	\begin{equation}
	 \mathcal{S}^{\rm eff} = 4\int \sigma m'dr + \int \frac{d}{dr}\Big[2\sigma'r(2m-r)+2\sigma(m'r-m)\Big]dr \ .
\label{ssigmam}
	\end{equation}
This EA is again of the form~\eqref{actionspatial3} with
	\begin{equation}\label{f2}
\mathcal{L}(\sigma, m; \sigma',m'; r)=4 \sigma  m' \ , \qquad	 f(\sigma, m; \sigma',m'; r) = 2\sigma'r(2m-r)+2\sigma(m'r-m) \ .
	\end{equation}
Again, admitting the existence of an event horizon, we take $r_i$ in~\eqref{actionspatial3} to be $r_i=r_H$, such that $2m(r_H)=r_H$ and applying~\eqref{virialea4} yields 
	\begin{equation}\label{TdVirialm}
	 \bigg[-2\sigma'(r^2+2mr_H-2rr_H)-2\sigma m' r_H\bigg]_{r_H}^\infty = 0 \ . 
	\end{equation}
For Schwarzschild, $m = M$ and $\sigma=1$, and this identity is trivially satisfied.

The peculiar feature of the (would be) virial identity~\eqref{TdVirialm} is the absence of the integral term; only the boundary term contributes. This is a consequence of the EH action for this ansatz being invariant (up to a boundary term) under the scaling transformation~\eqref{scaling1}, which is manifest from the fact that the integrand (plus integration measure) of the first term in~\eqref{ssigmam} is $\sigma \frac{dm}{dr}dr$. We learn, by example, therefore, that an appropriate choice of parameterization functions can simplify the virial identities by trivializing some terms. Thus, in spherical symmetry, the metric gauge~\eqref{metricans} with the $\sigma(r),m(r)$ parameterization functions~\eqref{mfunction} is the simplest choice for computing virial identities, which we shall therefore use in (most of) the following cases.

Again, we emphasise that, despite the check of the Schwarzschild solution (which now is more trivial), the (would be) virial identity~\eqref{TdVirialm} is \textit{incomplete}. The correct version will be given below in eq.~\eqref{virialvacuumok2}.

\subsection{Electrovacuum: an inconsistency}
\label{S43}
%
Our final example of this Section will make clear that there is one key ingredient missing in the computation of virial identities for GR. We now consider spherically symmetric solutions in electrovacuum. The action is~\eqref{totalaction} with 
\begin{equation}
\mathcal{S}_{\rm m}^{\rm Maxwell}=-\frac{1}{16\pi}\int d^4x \sqrt{-g} F_{\mu\nu}F^{\mu\nu} \ ,
\label{Maxaction}
\end{equation}
where $F_{\mu\nu}=\partial_\mu A_\nu-\partial_\nu A_\mu$ is the Maxwell field strength.  Following the conclusion at the end of the last subsection we take the  metric gauge~\eqref{metricans} with the $\sigma(r),m(r)$ parameterization functions~\eqref{mfunction}, and the ansatz for gauge potential 
\begin{equation}
A_\mu dx^\mu= -V(r)dt \ .
\label{electricansatz}
\end{equation}
Defining the EA as $\mathcal{S}_{\rm EH}+\mathcal{S}_{\rm m}^{\rm Maxwell}=(4\pi)^{-1}\int dt \mathcal{S}^{\rm eff}$, we find that the EA is again of the form~\eqref{actionspatial3} with
	\begin{equation}\label{f3}
\mathcal{L}(\sigma, m,V; \sigma',m',V'; r)=4 \sigma  m'+\frac{2r^2(V')^2}{\sigma} \ , \qquad	 f(\sigma, m; \sigma',m'; r) = 2\sigma'r(2m-r)+2\sigma(m'r-m) \ .
	\end{equation}
The difference with~\eqref{f2} is the extra term depending on $(V')^2$ in the effective Lagrangian. Applying~\eqref{virialea4}, the new  identity becomes
\begin{equation}\label{evvirial}
\int_{r_H}^\infty dr\frac{r (V')^2}{\sigma}(2r_H-r)=	 \bigg[-\sigma'(r^2+2mr_H-2rr_H)-\sigma m' r_H\bigg]_{r_H}^\infty  \ . 
	\end{equation}

If~eq.~\eqref{evvirial} were the correct virial identity, the RN solution, which has 
	\begin{equation}
\label{rnsol}
	 m(r) = M-\frac{Q^2}{2r} \ , \qquad \qquad \sigma = 1 \ , \qquad \qquad V(r)=-\frac{Q}{r} \ ,
	\end{equation}
should verify it. However, whereas the \textit{lhs} of~\eqref{evvirial} vanishes, the \textit{rhs} gives
	\begin{equation}\label{RNTD}
	 -  m' r_H\bigg|_{r_H}^\infty = \frac{Q^2}{2r_H}\neq 0 \ .
	\end{equation}
The fact that eq.~\eqref{evvirial}  is not satisfied for the RN solution means this is not the correct virial identity for the electrovacuum model. 
	
 In the next Section we propose that the boundary term of the gravitational action is mandatory in the correct treatment of  virial identities in GR.  This boundary term is the GHY term.  As we shall see, the contribution of such term for the vacuum case turns out to be trivial for the Schwarzschild solution with the parameterizations discussed in this Section. This explains the accidental (and thus misleading) check provided by the Schwarzschild solution to the incomplete vacuum GR virial identities~\eqref{VirialSphAns1} and~\eqref{TdVirialm}; but in the electrovacuum case, the boundary term provides a contribution to the incomplete virial identity~\eqref{evvirial} which is non-trivial for the RN solution and which precisely makes it verify the correct virial identity, given below in eq.~\eqref{evvirial2}.
%

\section{GR in spherical symmetry - adding the missing GHY term}
\label{Section5}
%
	The GHY~\cite{york1972role,gibbons1993action,hawking1996gravitational,brown1993microcanonical} term is a surface term that is necessary for GR to have a well posed variational principle in a manifold with a boundary. In the case of a BH spacetime (such as the Schwarzschild and the RN spacetimes), there are boundaries at  the horizon and at spatial infinity that, in principle, need to be considered. 

The complete gravitational action on a manifold $\mathcal{M}$, including the boundary term, is
	\begin{equation}
	 \mathcal{S}_{grav}= \mathcal{S}_{EH}+\mathcal{S}_{GHY} =\frac{1}{16\pi} \int_\mathcal{M} d^4x \sqrt{-g}R + \frac{1}{8\pi}\int_{\partial\mathcal{M}} d^3x \sqrt{-\gamma}(K-K_0)  \ ,
\label{ehghy}
	\end{equation}
	where $K =\nabla_\mu n^\mu$ is the extrinsic curvature of the boundary $\partial\mathcal{M}$ with normal $n^\mu$, and $\gamma$ is the associated 3-metric of the boundary. The extra $K_0$ term corresponds to the extrinsic curvature in flat spacetime
	(the background metric), necessary to obtain a finite result.

The GHY boundary term will give an extra total derivative to the EA. In this Section we will compute it  in the spherical case, under the parametrizations we have considered in Section~\ref{Section4}. This will remain consistent with the vacuum case and fix the issue raised in the electrovacuum case.

\subsection{Vacuum: $\sigma-N$ parameterization in Schwarzschild coordinates}
\label{S51}

	We consider again the metric ansatz \eqref{metricans}. Assume the spacetime has a boundary that is a spherical surface at a specific radius $r$ (like the spatial sections of the event horizon). Thus, the normal vector is $n = \sqrt{N}\partial_r$. Then 
	\begin{align}
	 \sqrt{-\gamma} &= \sigma\sqrt{N} r^2\sin\theta  \ ,\\
	 K&= \nabla_\mu n^\mu = \partial_r n^r + \frac{2}{r}n^r +\frac{\sigma'}{\sigma}n^r = \frac{1}{2}\frac{N'}{\sqrt N} + \left(\frac{2}{r}+\frac{\sigma'}{\sigma}\right)\sqrt{N} \ ,\\	
	 K_0&= \frac{2}{r} \ ,\\
	 \sqrt{-\gamma}(K-K_0) &= \left[\frac{r^2}{2}\sigma N'  + 2r\sigma( N-\sqrt{N}) +  r^2\sigma'N\right]\sin\theta \ .
	\end{align}

Defining as before an EA contribution for the GHY term, $ \mathcal{S}_{grav}=(4\pi)^{-1}\int dt \mathcal{S}^{\rm eff}$, we obtain an EA as in~\eqref{actionspatial3} with an \textit{extra} total derivative, defined by 
	\begin{equation}\label{fGHY}
	 f^{GHY} = r^2\sigma N' + 4r\sigma( N-\sqrt{N}) + 2r^2\sigma'N \ .
	\end{equation}
Comparing with~\eqref{f1}, the old $f$ cancels out completely. This removes the second derivatives from the complete EA (precisely the goal of the boundary term), which remains of the form~\eqref{actionspatial3} with
	\begin{equation}\label{f11}
\mathcal{L}(\sigma, N; \sigma',N'; r)=-2 \sigma \left(-1+N+rN'\right) \ , \qquad	 f(\sigma, N; \sigma',N'; r) =  4r\sigma( N-\sqrt{N})  \ .
	\end{equation}
 Then, the virial identity obtained from~\eqref{virialea4} is
	\begin{equation}\label{virialvacuumok}
	 2\int_{r_H}^{\infty}\sigma \left[N-1+(r-r_H)N'\right]dr= \bigg[4\sigma(N-\sqrt{N}) (r-r_H)\bigg]^{+\infty}_{r_H}  \ . \qquad  {\rm {\bf [Virial \  vacuum \ GR \ \sigma-N ]}}
	\end{equation}
This is the complete virial identity for vacuum GR in the $\sigma-N$ parameterization (correcting~\eqref{VirialSphAns1}). One can check that the Schwarzschild solution~\eqref{schwarzschildsol} still obeys it. The \textit{lhs} remains unchanged whereas the \textit{rhs} still gives $-4M$ (which now comes from the limit at $r=+\infty$).

\subsection{Vacuum: $\sigma-m$ parameterization in Schwarzschild coordinates}
\label{S52}

	For the $\sigma-m$ parameterization, on the other hand, where $N(r)$ is replaced by $m(r)$ via~\eqref{mfunction}, the extra total derivative from the GHY boundary term is
	\begin{align}
	 f^{GHY}  &= 2 r \sigma'(r-2 m)-2 \sigma \left[m'r+2r \sqrt{1-\frac{2 m}{r}}-2r+3 m\right] \ .
	\end{align}
Adding this contribution to the old $f$ in~\eqref{f2},  (again) cancels out the second derivatives in the complete EA which remains of the form~\eqref{actionspatial3} with
	\begin{equation}\label{f22}
\mathcal{L}(\sigma, m; \sigma',m'; r)=4 \sigma  m' \ , \qquad	 f(\sigma, m; \sigma',m'; r) = 
-4 \sigma \left[ \sqrt{r^2-{2 mr}}-r+2 m\right]  \ .
	\end{equation}
The virial identity obtained from~\eqref{virialea4} is then
	\begin{equation}\label{virialvacuumok2}
	 \bigg[-4\sigma\left(\frac{r-m}{\sqrt{r^2-2mr}}-1\right) (r-r_H)\bigg]^{+\infty}_{r_H}=0  \ . \qquad  {\rm {\bf [Virial \  vacuum \ GR \ \sigma-m ]}}
	\end{equation}
One can check that for the Schwarzschild solution ($\sigma=1$, $m=M$=constant) this is obeyed (considering carefully the $r=+\infty$ limit). Thus, this is the complete virial identity for vacuum GR in the $\sigma-m$ parameterization (correcting~\eqref{TdVirialm}).

\subsection{Electrovacuum: solving the inconsistency}
\label{S53}
From the results in Section~\ref{S43} and in the last subsection~\ref{S52} we can straightforwardly put together the virial identity for the electrovacuum case to be
\begin{equation}\label{evvirial2}
\int_{r_H}^\infty\frac{r (V')^2}{\sigma}(2r_H-r)=	 \bigg[-2\sigma\left(\frac{r-m}{\sqrt{r^2-2mr}}-1\right) (r-r_H)\bigg]^{+\infty}_{r_H} \ . 
\qquad  {\rm {\bf [Virial \   electrovacuum \ GR \ \sigma-m ]}}
	\end{equation}
It is now simple to check that the RN solution~\eqref{rnsol} verifies this virial identity (both \textit{lhs} and \textit{rhs} vanish).

%
\section{GR in spherical symmetry ($\sigma-m$ parameterization): illustrations}
\label{Section6}

Being in control of the correct methodology, we shall now compute the virial identity for different matter models. We shall always use the metric ansatz~\eqref{metricans} with the $\sigma-m$ parameterization~\eqref{mfunction}. The gravitational part of the action is given by $ \mathcal{S}_{grav}$, eq.~\eqref{ehghy}. This means the corresponding contribution to the virial identity is~\eqref{virialvacuumok2}. For the matter models to be considered here, this boundary term does not contribute. This is a consequence of the behaviour of $m(r)$ and $\sigma(r)$ at infinity and at the origin/horizon, depending on whether we consider solitonic solutions or BHs. At infinity these models have the asymptotic behaviour
\begin{equation}
\sigma(r)=1+\mathcal{O}\left(\frac{1}{r}\right) \ , \qquad m(r)=M+\mathcal{O}\left(\frac{1}{r}\right) \ , 
\end{equation}
and a careful analysis of the $r\rightarrow \infty$ limit of~\eqref{virialvacuumok2} shows it does not contribute. For the lower limit of~\eqref{virialvacuumok2}, the models we consider have the following behaviour close to the horizon
\begin{equation}
\sigma(r)=\sigma_H+\mathcal{O}\left(r-r_H\right)\ , \qquad m(r)=\frac{r_H}{2}+\mathcal{O}\left(r-r_H\right) \ ,
\end{equation}
and we can see that the limit will be proportional to $(r-r_H)^{1/2}$, rendering the horizon 
contribution zero; for solitons, at the origin,
\begin{equation}
\sigma(r)=\sigma_0+\mathcal{O}\left(r^{n_1}\right) \ , \qquad m(r)=\mathcal{O}\left({r^{n_2}}\right) \ , 
\end{equation}
where $n_1,n_2$ are model dependent but typically greater than 1
(for example, $n_2=3$ for all models discussed in this Section). 
This implies the $r=0$ contribution also vanishes.  
Thus, the whole contribution that one needs to consider to the virial identity comes from the matter action itself. 
This illustrates how the correct choice of parameterizing functions simplifies the computation of virial identities.

In all cases in this Section, we end up with an EA of the type~\eqref{actionspatial} with an effective Lagrangian
\begin{equation}
\mathcal{L}(\sigma,m,X;\sigma',m',X';r) \ ,
\end{equation}
where $X$ denotes collectively the parameterizing functions coming from the matter sector. The corresponding virial identity is then computed from~\eqref{virialea}.

For all models discussed in this Section, we have 
solved numerically the field equations and evaluated the displayed 
virial identities for a large sample of solutions in each case.
Although the relative errors depend on the values of various input parameters,
they are typical of order $10^{-5}$ or smaller.
An explicit illustration of this sort of numerical checking is provided in Section~\ref{S73}.

	\subsection{Solitonic solutions}
%
%
Let us start by considering solitonic solutions, thus without an event horizon. Therefore $r_i=r_H=0$.
%

		\subsubsection{Scalar boson stars}
\label{S611}
%
Scalar boson stars~\cite{Kaup:1968zz,Ruffini:1969qy} are self-gravitating lumps of a complex, massive scalar field - see also~\cite{schunck2003general,liebling2017dynamical,herdeiro2017asymptotically,Herdeiro:2020jzx}. They mimic $Q$-balls in their harmonic time-dependence. In spherical symmetry they are described by the same scalar field ansatz as $Q$-balls~\eqref{complexPhi}. But unlike the latter they do not require a  self-interacting scalar field; the necessary non-linearities are provided by GR.  

Consider the action that describes the self-gravitating complex scalar field, using the ansatz \eqref{complexPhi} in a model with a self-interactions potential $U(\Phi)$
			\begin{equation}\label{E88}
			 \mathcal{S}= \mathcal{S}_{grav}+\mathcal{S}_{\rm m}^{\Phi^*} \ ,	 
			\end{equation}
where the latter action is explicitly given by~\eqref{complexaction}. The resulting effective matter Lagrangian is,
\begin{equation}
 \mathcal{L} (\sigma, m, \phi; \sigma',m', \phi '; r) = r^2  \sigma \left[ \frac{r \omega ^2 \phi ^2}{(r-2 m) \sigma ^2}-\left(1-\frac{2 m}{r}\right) \phi'^{\,2} -U(|\phi|)\right] \ .
\end{equation}
Then, the virial identity reads
			\begin{equation}
			\int _0 ^{+\infty} dr \,  r ^2 \sigma \left[-\frac{r \omega^2 \phi ^2}{\sigma ^2}\frac{3r-8m}{(r-2m )^2}+\phi'^{\,2} +  3   \ U(|\phi|) \right] = 0  \ . \qquad {\rm {\bf [virial  \ scalar \ boson \ stars]}}
\label{virialbs}
			\end{equation}

For $m=0$, $\sigma=1$, this reduces to the $Q$-balls virial identity~\eqref{virialqballs}. Eq.~\eqref{virialbs} allows an immediate conclusion: if $\omega=0$ and the potential $U(\phi)$ is everywhere non-negative, the identity can never be respected, leading to a \textit{no-go} theorem~\cite{heusler1996no}. Thus gravity is not enough to circumvent Derrick's theorem; even with gravity, a finite oscillation frequency $\omega$ is necessary to have self-gravitating scalar solitons (with a time-independent spacetime). We will see in Section~\ref{S626} a distinct case: a matter model for which no solitons exist in flat spacetime but where the coupling to Einstein's gravity makes them possible.

		\subsubsection{Dirac stars}
%
	 Einstein's gravity minimally coupled with spin $1/2$ fields, allows the existence of self-gravitating solitons~\cite{finster1999particlelike}. These solitons are also known as \textit{Dirac Stars} - see also~\cite{dolan2015bound,herdeiro2017asymptotically,Herdeiro:2020jzx}. The corresponding action is 
	\begin{equation}
	 \mathcal{S} = \mathcal{S}_{grav}-\frac{i}{4\pi} \int d^4 x \sqrt{-g} \Bigg[ \frac{1}{2}\Big( \big\{ \hat{\slashed D} \overline{\psi}^{[A]}\big\}-\overline{\psi} ^{[A]} \hat{\slashed D} \psi ^{[A]} \Big)+U(\Psi)\Bigg]\ ,  
	\end{equation}
	where $\Psi $  is a Dirac $4$-spinor, with four complex components, while the index $[A]$ corresponds to the number of copies of the Lagrangian. For a spherically symmetric configuration one should consider, at least, two spinors with equal mass potential $U(\Psi)$; a single spinnor will necessarily make the solition rotate, yielding a stationary axially symmetric spacetime~\cite{Herdeiro:2019mbz}, rather than a spherical, static spacetime. The "dashed" derivative is $\hat{\slashed D}\equiv \gamma ^\mu \hat{D} _\mu$, where $\gamma ^\mu$ are the curved space gamma matrices and $\hat{D}=\partial _\mu +\Gamma _\mu $ is the spinorial covariant derivative, with $\Gamma _\mu $ being the spin connection matrices.
	
	For the Dirac field, the matter ansatz introduces two real functions $h(r)$ and $j(r)$
	\begin{eqnarray}
	 &&
	 \Psi^{[1]} = \begin{pmatrix} 
	 \cos(\frac{\theta}{2}) z(r)
	 \\ 
	 i \sin(\frac{\theta}{2}) \bar z(r)
	 \\ 
	 -i \cos(\frac{\theta}{2}) \bar z(r)
	 \\ 
	 -\sin(\frac{\theta}{2})  z(r)
	 \end{pmatrix}
	 e^{i(\frac{1}{2}\phi-\omega t) } \ , \qquad
	 \label{down}
	 \Psi^{[2]} = \begin{pmatrix} 
	 i \sin(\frac{\theta}{2}) z(r)
	 \\ 
	 \cos(\frac{\theta}{2}) \bar z(r)
	 \\ 
	 \sin(\frac{\theta}{2}) \bar z(r)
 	 \\ 
	 i \cos(\frac{\theta}{2})   z(r)
	 \end{pmatrix}
	 e^{i(-\frac{1}{2}\phi-\omega t) } \ ,
	\end{eqnarray}
	where $z(r) \equiv (1+i) h(r) +(1-i) j(r)$ and  $\Psi =i \bar{\psi}^{[A]}\psi^{[A]} = 4(h^2 - j^2 )$. The effective matter Lagrangian is
\begin{equation}
 \mathcal{L} (\sigma, m, h,j; \sigma',m', h',j'; r) = r^2 \sigma  \left[ \sqrt{1-\frac{2 m}{r}} \left(j h'-h j' \right)-\frac{\omega  \left(h ^2+j^2\right)}{\sqrt{1-\frac{2 m}{r}} \sigma }+\frac{2 hj }{r}+\frac{U(\Psi)}{4}\right]\ .
\end{equation}
Then, we get the virial identity\footnote{Here, and in some other cases below,  the identity is expressed in terms of $N$, rather than $m$, for compactness, although the computation is made with the  $\sigma-m$ parameterization.}
		\begin{equation}
		\label{d1}
	 \int _0 ^{+\infty} dr\ \frac{r^2 \sigma }{\sqrt{N} } \Bigg[(3 N+1) \left(j h'-h j'\right)+\frac{\omega  \left(h^2+j^2\right)}{\sigma }\left(\frac{1}{N}-7\right)+\left(\frac{ 8 hj}{r}+\frac{3}{2}  U(\Psi)\right) \sqrt{N} \Bigg]= 0 \ . \quad {\rm {\bf [virial \ Dirac \ stars]}}
		\end{equation}
 Differently from the scalar case,  
this identity does not provide any clear indication for the mechanism allowing the existence of solutions.
However, 
	in the flat spacetime limit, (\ref{d1}) reduces to
		\begin{equation}
	 \int _0 ^{+\infty} dr\ r^2 \Bigg[\big( j h'-hj'\big)+\frac{2 h j}{r}-\frac{3 }{2} \omega (h^2 +j ^2)+\frac{3}{8}U (\Psi)\Bigg] = 0 \ ,
		\end{equation}
	which can be further simplified through the field equations to yield
	\begin{equation}
	\int _0 ^{+\infty } dr \ r^2 U(\Psi) =\int _0 ^{+\infty} dr\ r^2 \Big[4\omega (h^2+j^2) \Big]\ .
	\end{equation}
Then, one observes that for a strictly positive potential, $U(\Psi)>0$,
the solutions are supported by the harmonic time-dependence, with $w>0$.

		\subsubsection{Vector boson stars (Proca stars)}
%
 Spherical vector boson stars, \textit{aka Proca Stars}~\cite{brito2016proca} (see also~\cite{SalazarLandea:2016fvo,Duarte:2016lig,herdeiro2017asymptotically,Minamitsuji:2017pdr,Minamitsuji:2018kof,Herdeiro:2020jzx}), can be found in GR minimally coupled to complex, massive vector fields. The model is described by the action
			\begin{equation}
			 \mathcal{S}= \mathcal{S}_{grav}-\frac{1}{4\pi} \int d^4 x \sqrt{-g} \Bigg[ F_{\mu \nu} F^{*\mu \nu} +V(\textbf{A})\Bigg]\ .
			\end{equation}
	where the complex vector field's ansatz is 
			\begin{equation}
			A_\mu =\big[ f(r) dt+i g(r) dr \big] e^{-i\omega t} \ ,
			\end{equation}
and $F_{\mu \nu} = \partial _\mu A_\nu -\partial _\nu A_\mu $. The vector field  is under a self-interacting potential $V(\textbf{A})$, where $\textbf{A}\equiv A_\mu A^{*\mu}$. One obtains the effective matter Lagrangian
\begin{equation}
\mathcal{L} (\sigma, m, g,f; \sigma',m', g',f'; r) = \frac{r^2}{\sigma}\left[ - \left(f'-\omega  g \right)^2 +\sigma ^2 V(\textbf{A})\right]\ .
\end{equation}
 The resulting virial identity is 
			\begin{equation}
			 \int _0 ^\infty dr \ \frac{r^2}{\sigma} \Bigg[-(\omega g - f')(3\omega g - f')+3 \sigma ^2 V(\textbf{A})+ \frac{1-N}{N^2} \frac{d V(\textbf{A})}{d\textbf{A}}(\sigma ^2 N^2 g^2 + f^2) \Bigg] =0\ . \qquad {\rm {\bf [virial \ Proca \ stars]}}
			\end{equation}
This identity reduces to the one in~\cite{brito2016proca} 
for a massive, free complex vector field. In the absence of self-interactions, the above relation can be used to rule out non-gravitating solutions.

\subsubsection{Einstein-Maxwell-Scalar (EMS) solitons} \label{PLEMS}
%
	The EMS model is described by the action
\begin{equation}\label{E113}
\mathcal{S} = \mathcal{S}_{grav}+ \frac{1}{4\pi} \int d^4 x \sqrt{-g} \Bigg[-\frac{1}{2}\partial _\mu \phi \partial ^\mu \phi - f(\phi ) F_{\mu\nu}F^{\mu \nu} - U(\phi )\Bigg] .
\end{equation}
In this model $F_{\mu \nu}$ is the Maxwell tensor and $\phi $ is a real scalar field that is non-minimally coupled to the Maxwell term through the coupling function $f(\phi) $. Moreover, we admit a self-interactions potential $U(\phi)$ for the scalar field. Particle-like soliton configurations were found in \cite{herdeiro2020class} (see also~\cite{herdeiro2019inexistence}). These configurations have a scalar field that depends only on the radial coordinate, $\phi \equiv \phi(r)$.

For an electric $4$-vector potential, $A_\mu = V(r) dt$, the resulting effective matter Lagrangian is
	\begin{equation}
	\mathcal{L} (\sigma, m, \phi ; \sigma',m', \phi '; r) =r^2 \sigma \left[f(\phi) \frac{2V'^{\,2}}{ \sigma ^2}- \left(1-\frac{2 m}{r}\right)   \phi '^{\,2} -U(\phi)\right]\ .
	\end{equation}
A first integral is obtained from the field equations, that simplifies the EA, namely,
\begin{equation}
V' (r) =-\frac{Q}{r^2 \varepsilon_\phi}\ ,
\end{equation}
	where $\varepsilon _\phi = f(\phi) \sigma ^{-1}$ can be thought as a relative electric permittivity that is caused by the non-minimal coupling between the scalar and Maxwell fields.	
	
	Replacing the first integral into the Maxwell term, the resulting virial identity is
\begin{equation}\label{PLEMSV}
	\int _0 ^{+\infty} dr \left[ r ^2 \sigma \phi'^{\,2} +3 r ^2 \sigma \ U(\phi ) - 2\frac{Q^2}{r^2 \varepsilon _\phi}  \right] = 0 \ . \qquad {\rm {\bf [virial \ EMS \ solitons]}}
\end{equation}
	The virial identity informs us that   particle-like solution can be supported by the electric charge or a negative potential.

%
	\subsection{Black holes}
%
As already mentioned in the Introduction, virial theorems can be used to establish no-hair theorems for BHs  (see~\cite{Herdeiro:2015waa} for a review). Heusler and Straumann obtained virial identities with that goal in \cite{heusler1996no} and \cite{heusler1992scaling} for the Einstein-Klein-Gordon model (that we shall refer to as \textit{scalar vacuum} - Section~\ref{Timedep}) and Einstein-Yang-Mills model (Section~\ref{S626}). 
In order to consider BHs, in this sub-Section we take $r_i=r_H\neq 0$.
%

		\subsubsection{No scalar hair theorem}
\label{Timedep}
%
The virial identity obtained for the model defined by~\eqref{E88} can be generalized to include a putative horizon scale $r_H$. Using a scalar field ansatz with a harmonic time-dependence~\eqref{complexPhi} one obtains\footnote{We remark that there is a factor of $1/2$ difference as compared to eq. (46) in~\cite{Herdeiro:2015waa}, which comes from a different action normalization.}
	\begin{align}\label{NHTEMS}
	 \int_{r_H}^\infty dr~&
	 \bigg\{\frac{1}{\sigma}\bigg[\frac{3(r-r_H)(r-2m)+r(r_H-2m)}{(r-2m)^2}\bigg]\omega^2r^2\phi^2\\ + \nonumber &
	 \sigma\left[\left( \frac{2r_H}{r}\left(1-\frac{m}{r}\right)-1
\right)r^2\phi'^2+\left(\frac{2r_H}{r}-3\right)r^2U(\phi)\right]
\bigg\}
=0 \ .  \qquad {\rm {\bf [virial \ scalar \ vacuum]}}
	\end{align}
Putting $r_H=0$ we recover~\eqref{virialbs}. On the other hand, putting $\omega=0$ one keeps only the second line. For this special case, inspection shows that the prefactor of $U$ and the first term (in the second line) are negative for $r>r_H$. This establishes a no-hair theorem for this model with $\omega=0$~\cite{heusler1996no}. This virial identity is not enough, however, to establish a no-hair theorem for $\omega\neq 0$, albeit such theorem can be established using other methods~\cite{pena1997collapsed,graham2014stationary}.

		\subsubsection{EMS BHs}\label{S321}
%
	Let us reconsider the EMS model \cite{herdeiro2018spontaneous,Fernandes:2019rez,Blazquez-Salcedo:2020nhs,LuisBlazquez-Salcedo:2020rqp} described by the action \eqref{E113}, but now taking into account the presence of an event horizon. Then, the virial identity reads
			\begin{equation}
			\int _{r_H} ^\infty dr\  \Bigg[I_\Phi (0, r_H)+ I_U ^{[\Phi]} (r_H)-f(\phi  ) I_M (r_H) \Bigg]\ =0 \  . \qquad {\rm {\bf [virial \ EMS \ BHs]}}
\label{virialemsbhs}
			\end{equation}
where the scalar terms are	
	\begin{eqnarray}
\label{vs1}
		I_\Phi(\omega,r_H) &=& \frac{1}{\sigma}\bigg[\frac{3(r-r_H)(r-2m)+r(r_H-2m)}{(r-2m)^2}\bigg]\omega^2r^2\phi^2 + 
		\sigma\left( \frac{2r_H}{r}\left(1-\frac{m}{r}\right)-1
		\right)r^2\phi'^2 \ , \\
\label{vs2}
		I_U ^{[\Phi ]} (r_H) &=& r U (2r_H-3 r)\sigma \ ,
	\end{eqnarray}
	whereas the Maxwell term reads
	\begin{equation}
	I_M (r_H) = 2\frac{ (2 r_H-r) Q^2 }{\varepsilon _\Phi ^2 r^3 \sigma } \ .
	\end{equation}
	As expected~\eqref{virialemsbhs} reduces to \eqref{PLEMSV} when $r_H=0$. The identity~\eqref{virialemsbhs}  tells us that a nontrivial scalar hair requires a nonzero electric charge. Indeed, as mentioned in Section~\ref{Timedep},  $I_\Phi (0, r_H)<0$ outside the horizon; furthermore (since $\sigma>0$) for a non-negative potential  $I_U ^{[\Phi]} (r_H)$ is non-positive outside the horizon; thus the positive contribution must come from the Maxwell term. Observe that when $Q=0$, and replacing $I_\Phi (0, r_H)\rightarrow I_\Phi (\omega, r_H)$, then~\eqref{virialemsbhs} becomes~\eqref{NHTEMS}.

		\subsubsection{Einstein-Maxwell-Vector (EMV) BHs}
%
	The EMV model \cite{Oliveira:2020dru,fan2016black} is described by the action
			\begin{equation}
			 \mathcal{S} = \mathcal{S}_{grav}+\frac{1}{4\pi} \int d^4 x \sqrt{-g} \left[-\frac{1}{4} G_{\mu \nu}G^{\mu \nu} - f(\textbf{B} ) F_{\mu\nu}F^{\mu \nu} - U(\textbf{B})\right] ,
			\end{equation}
	where $B_\mu $ is a real vector field that is non-minimally coupled to the Maxwell term $F_{\mu \nu} F^{\mu \nu}$ through the coupling function $f(\textbf{B}) $, for which self-interactions (and a mass term) are described by the potential $U(\textbf{B})$.
	For the vector field we consider, following \cite{Oliveira:2020dru}, a time-independent vector field  ansatz, $B_\mu dx^\mu = B_t (r) dt$. The vector field kinetic term is $G_{\mu \nu} = \partial _\mu B_\nu -\partial _\nu B_\mu$ and $\textbf{B}=B_\mu B^\mu$. Assuming a purely electric field, the effective matter Lagrangian becomes
	\begin{equation}
	 \mathcal{L} (\sigma, m, B_t ; \sigma',m', B_t '; r) = \frac{r^2}{\sigma} \left[ -B_t'^{\,2}-f(\textbf{B}) V'^{\,2}-\sigma ^2 U(\textbf{B})\right] \ .
	\end{equation}
Then, using  the electromagnetic equation of motion to obtain a first integral (the charge $Q$),
	\begin{equation}
	\nabla_\mu(f F^{\mu\nu}) = 0 \Rightarrow V' =- \frac{Q\sigma }{r^2 f} \ ,
	\end{equation}
the corresponding virial identity becomes
			\begin{equation}
			 \int _{r_H} ^{+ \infty} dr  \ \, \Bigg[ \frac{r-r_H}{r}\frac{N-1}{\sigma N^2} \frac{df(\textbf{B})}{d\textbf{B}} \frac{ Q^2 B_t ^2}{ r^2f(\textbf{B})^2 }+\frac{r(2r_H -r)}{\sigma}\left( B_t ^{'2}+\frac{Q^2 \sigma ^2}{r^4f(\textbf{B})}\right)  - I_U^{[B]} (r_H) \Bigg]=0\ , \ \  {\rm {\bf [virial \ EMV \ BHs]}}
			\end{equation}
	where $I_U ^{[B]}$ corresponds to the contribution from the potential of the vector field
	\begin{equation}
	I_U ^{[B]} (r_H) = r^2\sigma\Big[  3 U(\mathbf{B}) -\frac{r-r_H}{r}\frac{N-1}{\sigma N^2}\frac{dU(\mathbf{B})}{d\mathbf{B}}B_t^2\Big] \ .
	\end{equation}
For flat spacetime and $U(\mathbf{B})=0$ this reduces to
	\begin{equation}\label{VIdFlat}
	\int_{0}^\infty dr\frac{1}{r^2} \left(r^4 B_t'^2+\frac{Q^2}{f(\textbf{B})}\right) = 0 \ .
	\end{equation}

If $f>0$, the virial identity~\eqref{VIdFlat} informs us that only the trivial configuration $B'_t=0$ and $Q=0$ is possible. In this case, of course, $B_\mu$ also became a gauge field (since the mass term vanished).

	\subsubsection{Einstein-Yang-Mills (EYM) BHs and solitons}
\label{S626}
%
	Yang-Mills theories \cite{yang1954conservation} are gauge theories based on non-Abelian Lie groups. These theories are at the core of the standard model of particle physics. Minimally coupling these "matter" models to Einstein's gravity leads to 
	EYM theories, which are described by the action
	\begin{equation}
	\mathcal{S}=\mathcal{S}_{grav}-\frac{1}{8\pi} \int d^4 x\  \sqrt{-g} {\rm Tr}(F^2)\ .
	\end{equation}
	As an illustration of the role of virial identities in EYM models, let us follow the work done by Heusler~\cite{heusler1996no}. One considers the purely magnetic $SU(2)$ configuration with the gauge potential $1-$form $A$ 
	\begin{equation}
	 A=[p(r)-1](\tau _\varphi d\theta -\tau_\theta\sin \theta d\varphi )\ .
	\end{equation}
The usual basis of $SU(2)$ is denoted as $(\tau_r,\tau_\theta,\tau_\varphi)$~\cite{Bartnik:1988am}; also $\tau _\theta \equiv \partial _\theta \tau _r $, $\tau _\varphi \sin \theta \equiv \partial _\varphi \tau _r$ and $\tau _r \equiv (2 i |\overrightarrow{r}|)^{-1} \overrightarrow{r}\cdot \overrightarrow{\delta}$; $p(r)$ is an unkown radial function, determined by solving the field equations. 
The effective matter Lagrangian is
	\begin{equation}
	 \mathcal{L} (\sigma, m, p ; \sigma',m', p'; r) = \sigma \left[\frac{1}{2}\left(1-\frac{2m}{r}\right)p'^{\,2} +\frac{(1-p^2)^2}{4r^2}\right]\ .
	\end{equation}
The virial identity in the presence of an event horizon is
	\begin{equation}
	 \int _{r_H} ^{+\infty} dr \ I_{YM} (r_H) =0\ , \qquad\qquad {\rm {\bf [Virial \ EYM]}}
	\end{equation}
	where the Yang-Mills term is
	\begin{equation}
	 I_{YM}(r_H) = \frac{\sigma}{2} \left\{\left[1+\frac{2 m}{r}\left(\frac{r_H}{r}-2\right)\right]N p'^2+\left[1-\frac{2r_H}{r}\frac{(1-p^2)^2}{2r^2}\right]\right\}\ .
	\end{equation}
	In the presence of a horizon, the virial identity does not exclude the existence of BHs with hair. In fact these BHs exist~\cite{Volkov:1989fi,Volkov:1990sva,Bizon:1990sr,1990JMP....31..928K} and were an influential counter-example to the no-hair conjecture~\cite{Bizon:1994dh,Volkov:1998cc}. The same occurs when $r_H\rightarrow 0$: the virial identity allows the existence of self-gravitating solitonic objects. In fact these solitons exist, as first pointed out by Barnik and Mckinnon~\cite{Bartnik:1988am}. However, in the absence gravity
	\begin{equation}
	 \int _0 ^{+\infty} dr \Bigg[ \frac{p'^2}{2}+\frac{(1-p^2)^2}{4r^2}\Bigg] =0 \ ,
	\end{equation}
	which shows that no flat spacetime Yang-Mills solitons exist. So, in this case, the coupling of the Yang-Mills source to Einstein's gravity is enough to allow particle-like solutions, which are forbidden in flat spacetime.

%
\subsubsection{Einstein-Maxwell-gauged scalar (EMgS) BHs}
%
	A \textit{gauged} complex scalar field minimally coupled to both the electromagnetic field and Einstein's gravity is described by the action 
	\begin{equation}
	\mathcal{S} = \mathcal{S}_{grav}+\frac{1}{4\pi} \int d^4 x\sqrt{-g} \Bigg[- \frac{1}{4}F_{\mu \nu } F^{\mu \nu} -g^{\mu\nu} D_{(\mu} \Phi D_{\nu)}^* \Phi^* - U(|\Phi|)\Bigg] \ ,
	\end{equation}
 	where $D_\mu = \partial_\mu -ieA_\mu$ is the covariant gauge derivative. In this case the global $\textbf{U}(1)$ symmetry of the scalar field is gauged. Charged (gauged) boson stars in this model have been discussed in~\cite{Jetzer:1989av,Pugliese:2013gsa}. 
	Hairy BHs in this class of models (with self-interactions) are also 
	possible and have been discussed in~\cite{Herdeiro:2020xmb,Hong:2020miv}.

 	For a purely electric spherical configuration~\eqref{electricansatz} and a scalar field with a harmonic time-dependence~\eqref{complexPhi}, we get
 the following effective matter Lagrangian
 		\begin{eqnarray}
 			\mathcal{L} (\sigma, m, \phi, V; \sigma',m', \phi ',V'; r) =
			r^2  \sigma \left[
			\left(1-\frac{2 m}{r}\right) \phi'^2
			 +U(|\phi|)
			-\frac{ (\omega-e V)^2 \phi^2}{(1-\frac{2m}{r})\sigma^2}
			-\frac{V'^2}{2\sigma^2}
			\right]~.
 		\end{eqnarray}
Then the
 corresponding virial identity for BH solutions reads \cite{Herdeiro:2020xmb}
\begin{eqnarray}
&&
\label{virialgsM}
\int_{r_H}^\infty dr~ r^2 \sigma
\left\{
     \left[
		             1-\frac{2r_H}{r}\left(1-\frac{m}{r}\right) 
		 \right]\phi'^2
+ \left(3-\frac{2r_H}{r} \right)U(|\phi|)
\right\}
\\
\nonumber
&&
=
\int_{r_H}^\infty dr~ r^2 
\left\{
\left(1-\frac{2r_H}{r}\right)\frac{V'^2}{2\sigma}
+
\left[
3-\frac{2r_H}{r}\left(1-\frac{3m}{r}\right)-\frac{8m}{r}
\right]\frac{(\omega-e V)^2 \phi^2}{N^2 \sigma}
\right\}~,  \quad\qquad {\rm {\bf [Virial \ EMgS]}}	
\end{eqnarray}
which reduces to (\ref{NHTEMS}) for $e=V=0$ case.
One notices that 
 both factors in front of
the  scalar quantities on the $lhs$  have a fixed, positive sign, such that all this integral is strictly positive
(here we assume $U(|\phi|)>0$).
Therefore no solutions with $\phi \neq 0$ can exist
 for $V=0$ (no Maxwell field) and $\omega=0$. 
Also, the factors in front of the  Maxwell quantities
on the $rhs$ 
 are indefinite (although they become positive asymptotically).
 Thus, for $V\neq 0$ and/or $\omega\neq 0$ a solution becomes possible (but not guaranteed). 

\section{GR in spherical symmetry and isotropic coordinates}
\label{Section7}
%
	An alternative coordinate system to deal with spherical spacetimes, often useful, is given by \textit{isotropic} coordinates - see $e.g.$~\cite{Townsend:1997ku}. In isotropic coordinates the radial coordinate is not the areal radius. In this Section we shall compute the virial identity in isotropic coordinates for two cases: electrovacuum and (massive, complex) scalar vacuum. We shall see that the correct virial identities, that include a non-trivial contribution from the GHY boundary term, are obeyed by known solutions of these models (the RN BH and boson stars). This gives us a further confirmation that  the GHY term is indeed required to construct the virial identity in a generic coordinate system and parameterization.
	
	\subsection{A general result}
\label{S71}
	Let us consider a general model, described by the action 
	$\mathcal{S}= \mathcal{S}_{grav}+\mathcal{S}_{\rm m}$,
	where 
	$\mathcal{S}_{grav}$ 
	includes also the GHY boundary term,
	while 
	$\mathcal{S}_{\rm m}$
	is the matter field(s) action
		(with the  presence of first order derivatives, only).
 As for the line element, 
we consider a general form  in terms of two functions $f_0,f_1$
\begin{eqnarray}
\label{metric-iso}
 ds^2=-f_0^2(r) dt^2+
f_1^2(r)\left[dr^2+r^2 (d\theta^2+\sin^2\theta d\varphi^2)\right] \ .
\end{eqnarray}
 
The computation of the gravity effective action 
is very similar to the case of Schwarzschild coordinates.
Although
the bulk action
$R\sqrt{-g}$ depends again on the second derivatives
of the metric functions $f_0,f_1$,
they can be collected into a \textit{total derivative}, 
such that this EA is cast in the form~\eqref{actionspatial3} with
\begin{eqnarray}
\label{iso1}
	\int dr  f_0f_1^3 r^2 R  =
		\int dr 
	\left[
 2r^2 \left(2f_0'f_1'+\frac{f_0f_1'^2}{f_1}\right)+
 \frac{df }{dr} 
\right] \  ,~~{\rm with}~~ f=-2r^2(f_1 f_0'+2 f_0 f_1')~.
\end{eqnarray}
We assume again that the 
 spacetime boundary is a spherical surface
at some radius $r$, with a normal vector $n=1/f_1\partial_r$.
Then one finds\footnote{Note that, in computing $K_0$,
one considers a (flat) background metric with a two sphere of radius $r f_1$.
}
	\begin{align}
	 \sqrt{-\gamma} &=  f_0f_1^2 r^2\sin\theta  \ ,
	\\
	 K&= \nabla_\mu n^\mu =\frac{1}{f_1}\left(\frac{2}{r}+\frac{f_0'}{f_0}\right)+\frac{2f_1'}{f_1^2} \ ,
	\\	
	 K_0&= \frac{2}{r f_1} \ ,
	\\
	 \sqrt{-\gamma}(K-K_0) & =  r^2( f_1 f_0'+2f_0f_1' ) \sin\theta \ .
	\end{align}
One can easily see that,
different from the case of Schwarzschild-like coordinates,
 the contribution of the GHY boundary term
cancels out {\it completely}  
the total derivative in the gravity bulk action (\ref{iso1}).
Then one finds the following gravity effective 
Lagrangian
 \begin{equation}
 \mathcal{L} (f_0,f_1; f_0',f_1'; r) =  2r^2 (2f_0'f_1'+\frac{f_0f_1'^2}{f_1}).
\end{equation}

When adding the EA for the matter sector of the model, 
the result  \eqref{virialea4}
implies the following 
 form of the
generic
 virial identity  
\begin{eqnarray}
\label{iso2}
{\cal V}_g+{\cal V}_m =0 \ ,   \qquad \qquad {\rm {\bf [virial \ isotropic \ general]}}
\end{eqnarray} 
with the  gravity contribution
\begin{eqnarray}
\label{iso3}
{\cal V}_{g}=
-2\int_{r_i}^\infty dr
\left[ 
 r(r-r_i)f_1 
\left(
2f_0'+\frac{f_0 f_1' }{f_1}
\right)
\right] ,
\end{eqnarray}
${\cal V}_m $
being the matter contribution
 (as resulting from 
 \eqref{virialea4},
in terms of matter field(s) effective Lagrangian $ \mathcal{L}_m$).

	\subsection{Electrovacuum}
\label{S72}
%
As the simplest application of the above results, 	
let us consider the   
 electrovacuum case,
with the Maxwell action as given by~\eqref{Maxaction}.
 The electric field is again purely electric, with
$A_\mu dx^\mu=V(r)dt$,
while
the Maxwell equations can be integrated to give
\begin{equation}
V'(r)=\frac{Q}{r^2}\frac{f_0}{f_1} \ ,
\label{chargeiso}
\end{equation}
with $Q$ the electric charge.
	
The contribution
${\cal V}_m$
 of the Maxwell field to the virial
(\ref{iso2})
 is computed from~\eqref{virialea4}
(with  $\mathcal{L}_M=2r^2 f_1 V'^2/f_0$).
After using (\ref{chargeiso})
  the final result
	reads
\begin{equation}
\int_{r_H}^\infty dr\left\{ 
\frac{f_0 Q^2}{f_1 r^3}+rf_1'\left(
2f_0'+\frac{f_0f_1'}{f_1}
\right)
\right\}(r-2r_H)=0 \ .   \qquad {\rm {\bf [virial \ electrovacuum \ isotropic]}}
\label{isotropicv1}
\end{equation}

After replacing the expression of the RN solution
	\begin{equation}
	f_0(r)=\frac{1-\frac{r_H^2}{r^2}}{1+\frac{M}{r}+\frac{r_H^2}{r^2}}\ , \qquad  
	f_1 (r)=1+\frac{M}{r}+\frac{r_H^2}{r^2},  \qquad {\rm where} \qquad r_H^2=\frac{M^2 -Q^2}{4}\ ,
	\end{equation}
the identity~\eqref{isotropicv1} simplifies to
\begin{equation}
\int_{r_H}^\infty dr
\frac{4r_H^2}{r^3}(r-2r_H)=4r_H^2\left( \frac{r_H}{r^2}-\frac{1}{r}\right)\bigg|^\infty_{r_H}=0 \ .  
\end{equation}
This confirms the RN solution obeys the identity~\eqref{isotropicv1}. 
Had we not included the GHY contribution, however, there would be an extra contribution to the identity coming from 
$f=-4r^2f_0f_1'-2r^2f_1f_0' $ in~\eqref{iso1}. 
Then, from~\eqref{virialea4}, 
this would give the extra contribution to the virial identity (\ref{iso2})
\begin{equation}
\left[\frac{\partial f}{\partial r}(r-r_i) - \sum_i\frac{\partial f}{\partial q'_i} q'_i\right]^{+\infty}_{r_H}=-\left[2r(r-2r_H)(2f_0f_1'+f_1f_0')\right]^{+\infty}_{r_H}=2(M-2r_H) \ .
\end{equation}
The fact that this is  non-vanishing  for $Q\neq 0$
  means that a virial identity derived solely from the EH plus 
Maxwell actions is not obeyed by the RN solution 
(albeit, accidentally, 
it is obeyed by the Schwarzschild solution as in the discussion of Section~\ref{Section4}). 
The correct identity must be derived from the full gravitational action,
 including the GHY boundary term.
 Moreover, using isotropic coordinates the contribution 
of the gravitational action to~\eqref{isotropicv1}
 is non-vanishing (and both the EH and GHY terms must be considered) 
unlike the special "gauge" discussed in Section~\ref{Section5}.

%
	\subsection{(Massive-complex) scalar vacuum}
\label{S73}
%
As a second illustration, let us reconsider the scalar boson stars already discussed in Section~\ref{S611}.
 The action is given by~\eqref{E88} and the scalar field ansatz is given by~\eqref{complexPhi}. 
Here, in order to test the virial identity for concrete solutions, 
we take the simplest choice for the potential, with 
a mass term only, $U(|\phi|) = {\mu^2} \phi ^2$.
Employing again the metric ansatz  \eqref{metric-iso}
this results in the scalar field effective Lagrangian
 \begin{eqnarray}
 \mathcal{L}_s= r^2 f_0f_1^3 \left[ \frac{\phi'^2}{f_1^2}+\left(\mu^2 -\frac{w^2}{f_0^2}\right)\phi^2 \right] \ .
 \end{eqnarray}
In the absence of an event horizon, the scaling of the radial coordinate is simply $r\rightarrow \tilde{r}= \lambda r$. 
Then, following the standard procedure,
 we obtain the simple expression for the scalar field contribution to the  virial identity
 (\ref{iso2})
	\begin{eqnarray}
	\label{E202}
	{\cal V}_m= 4\int _{0} ^{+\infty} dr ~r^2 f_0f_1 
	                \left[
	  \phi'^2+3f_1^2\left(\mu^2-\frac{w^2}{f_0^2}\right)\phi^2
	                \right] \ .
	\end{eqnarray}
Then, the whole virial identity~(\ref{iso2}) reads
\begin{equation}
\int_{0}^\infty dr
\left\{ 
 r^2f_1 
\left(
2f_0'+\frac{f_0 f_1' }{f_1}-2 f_0  \left[  \phi'^2+3f_1^2\left(\mu^2-\frac{w^2}{f_0^2}\right)\phi^2
	                \right]
\right)
\right\}
=0 \ .   
\qquad {\rm {\bf [virial \ boson \ stars \ isotropic]}}
\end{equation}

\begin{figure}[h!]
\begin{center}
\includegraphics[width=0.495\textwidth]{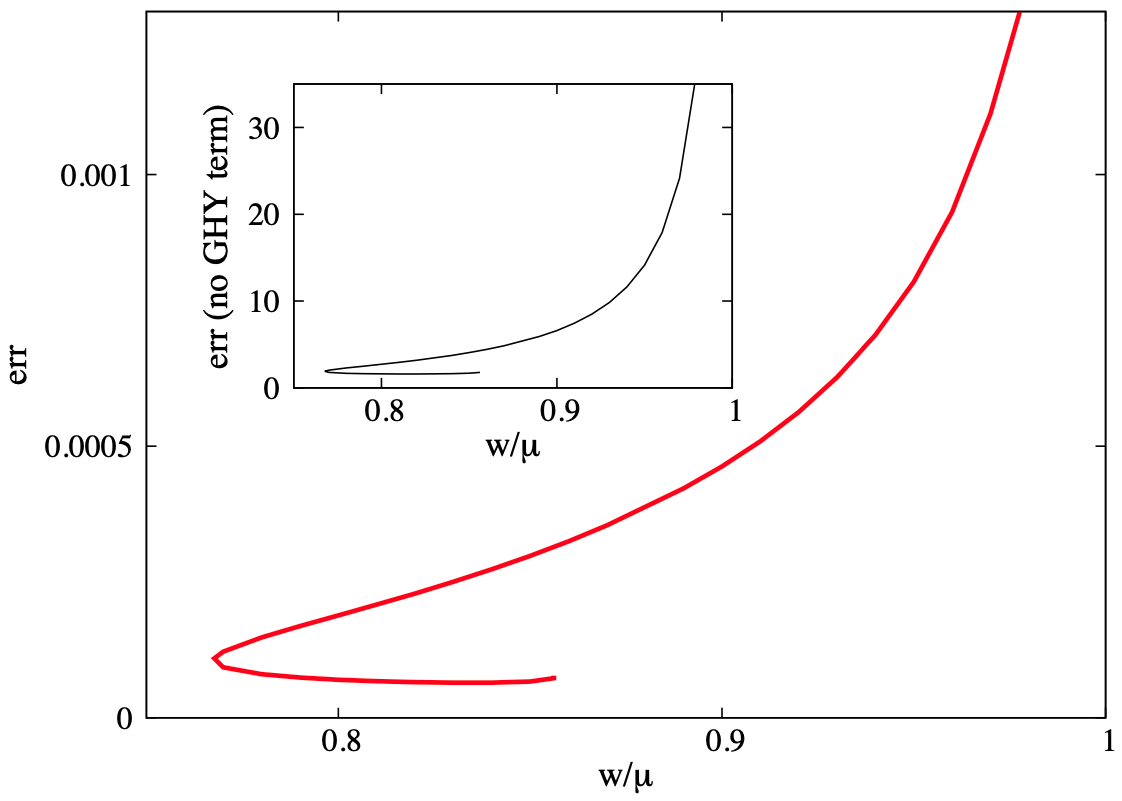} 
\caption{\small{ 
The relative error  (\ref{error})
for the virial identity satisfied by numerical boson stars in isotropic coordinates
is shown as a function of the 
ratio between the field frequency and field's mass.
The inset shows the same relative error but without including the boundary term in ${\cal V}_g$.
}}
\label{err}
\end{center}
\end{figure}

Differently from the electrovacuum case, no exact solutions are known for a boson star. Thus, to check numerically the validity of the relation
 (\ref{iso2}),
we define a relative error
	\begin{eqnarray}
	\label{error}
 err=1+\frac{{\cal V}_g}{{\cal V}_m},
	\end{eqnarray}
which would vanish for an infinity accuracy solution.
However, as seen in Fig.~\ref{err},
 $err$ is never zero for a numerical solution\footnote{In constructing the
boson stars in isotropic coordinates, we have used the approach described in 
Ref. \cite{Herdeiro:2015gia}
 (and in particular the same solver and the same grid choice).
The increase of $err$ as $w\to \mu$
can be attributed to the delocalization of the solutions in this limit,
with $\phi\to 0$ and $(f_1,f_0)\to 1$.

 }, 
and takes values compatible  with other  error estimates.  
The natural interpretation of this result is that
the virial relation  (\ref{iso2})
holds also for boson stars in isotropic coordinates.
 
	As for the role of the GHY term, an analogous computation to the one of the previous subsection  yields
	(taking into account the asymptotic behaviour 
	of the boson stars)
 an extra  $-2M$ contribution to the gravity part in the virial identity
	(with $M$ the ADM mass).
This is fundamental for the solutions to obey the virial identity. 
In Fig. \ref{err} (inset), we show the same relative error as in the main panel, but where ${\cal V}_g$ does not include the contribution from the GHY boundary term. One observes the error becomes order unity or larger, in this case.
  
\section{Conclusions and discussion}
\label{Section8}

To goal of this paper is to present a primer for a clear and efficient understanding of virial identities in non-linear field theories, in particular in relativistic gravity. As explained in  Section~\ref{Section2}, virial identities result from a specific type of variational principle obtained from an EA. Thus, they should be obeyed by the solutions of the Euler-Lagrange equations obtained from that EA, which extremize \textit{any} variation. Nonetheless, virial identities are integral identities that \textit{appear} independent from the field equations. Thus, their analysis provides different insights and checks than the ones provided by the analysis of the (differential) field equations.

In spherical symmetry, considering an appropriate ansatz in any non-linear field theory leads to an EA in the radial variable. Then, eqs.~\eqref{virialea}, \eqref{virialea2}, \eqref{virialea3} and \eqref{virialea4} provide a straighforward way to compute the virial identity. But it is mandatory that the EA contains all terms necessary to completely define the model. In the case of non-linear field theories for which the original action contains second derivatives of the fundamental variables, the well-posedness of the field equations in manifolds with boundaries requires the introduction of boundary terms. Whereas the latter are irrelevant for many analyses (such as computing the bulk solutions of the field equations), such boundary terms can, and in general will, contribute to virial identities. This is the case of GR, for which the EH action has second order derivatives of the metric and the complete gravitational action~\eqref{ehghy} needs the GHY boundary term. We have shown that this term must be considered in order to derive the correct virial identities in GR.

Nonetheless, there is a special "gauge" choice (corresponding to the $\sigma-m$ parameterization in Schwarzschild coordinates~\eqref{metricans} with~\eqref{mfunction}) where one can get away with neglecting the boundary term and indeed the whole gravitational action for the virial identity. This is because the EH action for this "gauge" choice leads to a scale invariant EA and the GHY boundary term does not contribute, at least for the boundary conditions that  apply to asymptotically flat regular solitons or BHs. In this context, it is important to stress that the scaling transformation leading to virial identities is \textit{not} a diffeomorphism; the EA results from the integral of scaled configurations which is not simply a coordinate transformation in the integral. Thus, in general, the EH action will contribute to virial identities. But it turns out that there is a nice "gauge" choice for which it does not, facilitating thus the computation of virial identities. 

This paper was focused on 1D EAs that are applicable to spherical configurations. Having understood clearly the foundations of the method we shall consider $n$D EAs and the particular example of axially symmetric configurations in GR in a companion paper~\cite{companion}. Another interesting question, that we hope to consider in the future, is the case of modified gravity, for which the boundary term needs to be appropriately modified.

%
\section*{Acknowledgements}
%
This work is supported by the Center for Research and Development in Mathematics and Applications (CIDMA) and by the Center for Astrophysics and Gravitation (CENTRA) through the Portuguese Foundation for Science and Technology (FCT - Funda\c c\~ao para a Ci\^encia e a Tecnologia), references UIDB/04106/2020, UIDP/04106/2020 and UIDB/00099/2020  and by national funds (OE), through FCT, I.P., in the scope of the framework contract foreseen in the numbers 4, 5 and 6 of the article 23, of the Decree-Law 57/2016, of August 29, changed by Law 57/2017, of July 19.
J. Oliveira  is supported by an FCT post-doctoral grant through the project PTDC/FIS-OUT/28407/2017 and A. Pombo is supported by the FCT grant PD/BD/142842/2018.  We acknowledge support from the projects PTDC/FIS-OUT/28407/2017,  CERN/FIS-PAR/0027/2019 and PTDC/FIS-AST/3041/2020. This work has further been supported by the European Union's Horizon 2020 research and innovation (RISE) programme H2020-MSCA-RISE-2017 Grant No.~FunFiCO-777740. The authors would like to acknowledge networking support by the COST Action CA16104. 

\appendix 
\addcontentsline{toc}{section}{APPENDICES}

\section{Derrick's theorem in higher dimensions}\label{AppendixA}
Consider the $D=n+1$ dimensional flat spacetime with the metric
\begin{equation}
ds_D^2 = -dt^2 + \sum_{i}^{n}dx_i^2 \ .
\end{equation}
The scalar field action is now
\begin{equation}
\mathcal{S}^D=\int dt \int d^{n}{\bf r}\left[-\partial_M \Phi \partial^M \Phi -U(\Phi)\right] \ .
\end{equation}
where the index $M$ takes values between $0$ and $n$. By following the same arguments as above, we obtain
\begin{equation}
\mathcal{S}^D=-\int dt E^D = -\int dt (I_1^D + I_2^D) \ ,
\end{equation}
where
\begin{equation}
I_1^D\equiv \int d^{n}{\bf r} (\nabla_{n} \Phi)^2 \ , \qquad I_2^D\equiv \int d^{n}{\bf r} U(\Phi) \ ,
\end{equation}
with $\nabla_{n}$ being the $n$ dimensional spatial gradient. Assuming once again the same 1-parameter family of configurations $\Phi_\lambda({\bf r})=\Phi(\lambda {\bf r})$ and extremizing the energy in the same way, we obtain the following virial identity
\begin{align}
\left(\frac{dE^D_\lambda}{d\lambda}\right)_{\lambda=1}&= (-n+1)I_1^D-(n+1)I_2^D =0 \ .\qquad {\rm {\bf [virial \ Derrick \ higher \ D]}}\label{virialD}
\end{align}
Moreover, the stability condition is, using the virial identity,
\begin{align}
\left(\frac{d^2E^D_\lambda}{d\lambda^2}\right)_{\lambda=1}&= n(-n+1)I_1^D -(n+1)(-n-2)I_2^D = 2(-n+1)I_1^D \ .
\end{align}
We see that for any $n>1$, we always have that any solution to the Klein-Gordon equation is unstable. At the same time, the virial identity \eqref{virialD} shows that both of the terms involved have the same sign for $n>1$ and a positive definite potential, meaning that, in such case, there are no solutions regardless of stability.
%

%

%
  \bibliographystyle{ieeetr}
  \bibliography{biblio}


\end{document}